\documentclass[aps,twocolumn,prd,showpacs,preprintnumbers,nofootinbib]{revtex4}
\usepackage{amsmath}
\usepackage{graphicx}
\usepackage{subfigure}
\usepackage{dcolumn}
\usepackage{bm}
\usepackage{amssymb}
\usepackage{latexsym}
\usepackage{psfrag}

\newcommand{\setC}{\mathbb{C}}

\newcommand{\setR}{\mathbb{R}}

\newcommand{\ie}{\emph{i.e.~}}
\newcommand{\GReCO}{${\cal G}\setR\varepsilon\setC{\cal O}$}

\newcommand{\df}{\delta\varphi}

\newcommand{\mean}[1]{\left\langle #1 \right\rangle}
\newcommand{\mpl}{m_{_\mathrm{Pl}}}
\newcommand{\MPl}{M_{_\mathrm{Pl}}}
\newcommand{\ini}{\mathrm{in}}\newcommand{\cl}{\mathrm{cl}}
\newcommand{\tin}{{t_\mathrm{in}}}
\newcommand{\phiin}{{\varphi_\mathrm{in}}}
\newcommand{\dd}{\mathrm{d}}

\renewcommand{\t}{\tau}
\newcommand{\e}{\mathrm{e}}
\newcommand{\s}{\sigma}

\newcommand{\tr}{\mathrm{Tr}}

\begin{document}

\preprint{UTTG-14-05}

\title{Solving Stochastic Inflation for Arbitrary Potentials}

\author{J\'er\^ome Martin} \email{jmartin@iap.fr}
\affiliation{Institut d'Astrophysique de Paris, \GReCO, UMR 7095-CNRS,
Universit\'e Pierre et Marie Curie, 98bis boulevard Arago, 75014
Paris, France}

\author{Marcello Musso} \email{musso@physics.utexas.edu}
\affiliation{University of Texas at Austin, Department of
Physics - Theory group, 1 University Station C1608, Austin TX
78712-0269 USA}

\date{\today}

\begin{abstract}
A perturbative method for solving the Langevin equation of
inflationary cosmology in presence of backreaction is presented. In
the Gaussian approximation, the method permits an explicit calculation
of the probability distribution of the inflaton field for an arbitrary
potential, with or without the volume effects taken into account. The
perturbative method is then applied to various concrete models namely
large field, small field, hybrid and running mass inflation. New
results on the stochastic behavior of the inflaton field in those
models are obtained. In particular, it is confirmed that the
stochastic effects can be important in new inflation while it is
demonstrated they are negligible in (vacuum dominated) hybrid
inflation. The case of stochastic running mass inflation is discussed
in some details and it is argued that quantum effects blur the
distinction between the four classical versions of this model. It is
also shown that the self-reproducing regime is likely to be important
in this case.
\end{abstract}


\maketitle


\section{Introduction}

Quantum effects play a crucial role during inflation. In particular,
they are responsible for the self-reproducing behavior of the Universe
also known as ``eternal inflation''~\cite{early}. In this regime, the
quantum fluctuations are so important that they can dominate the
classical dynamics and, therefore, treating them properly becomes
mandatory. Technically, this is a difficult task especially when it is
necessary to take into account the backreaction of the quantum field
on the geometry. Stochastic inflation~\cite{early, stocha,
rey,NNS,Kandrup,NS,N} aims at providing a formalism where the previous
difficulties can be partially circumvented. In the stochastic
inflation approach, one is mainly interested in the evolution of a
coarse-grained field, typically the original scalar field averaged
over a Hubble patch, and the quantum effects are modeled by a
stochastic noise originating from the small-scale Fourier
modes. Consequently, the dynamics of the coarse-grained field is
controlled by a Langevin equation. Then, endowed with a solution of
this equation, one can compute the probability density function of the
field and the various correlation functions.

\par

Even if stochastic inflation simplifies the calculation of the quantum
effects, the Langevin equation remains difficult to solve (without, of
course, relying on numerical computations). The case without
backreaction (in a de Sitter background) has been investigated in
Ref.~\cite{SY} where it has been shown that solutions for an arbitrary
potential can be obtained. The case with backreaction is clearly much
more complicated and something can be said about the solution only for
very specific potentials. The usual approach is applicable when the
inflaton potential is such that an exact solution exists for some
power of the field. Then, in order to obtain the stochastic field
itself and its various correlation functions, an expansion in terms of
the coupling constant of the potential is performed. Explicit examples
of this approach can be, for instance, found in
Refs.~\cite{Hodges,YVM} and more recently in Ref.~\cite{GT} and are
briefly discussed in the following. The main point is that it is first
necessary to obtain a solution to be subsequently able to perform the
expansion. This is the reason why the applicability of the method is
severely limited.

\par

In this article, we present a method based on a perturbative expansion
in the stochastic noise, the expansion being performed directly in the
Langevin equation. As a consequence, our method does not require
getting first a solution and, {\it a priori}, can be pushed to any
order, the only limitation being the mathematical complexity of the
obtained expressions. This represents a crucial advantage over the
other approaches which allows us to treat analytically the case of an
arbitrary potential with backreaction. It should be noticed that the
idea to solve perturbatively the Langevin equation was put forward for
the first time in Ref.~\cite{GLMM}. In that reference, the method was
used to compute the three-point correlation function of the Cosmic
Microwave Background (CMB) anisotropy while, here, we use it in order
to study the influence of the quantum effects on the behavior of the
background inflaton. In particular, we show that, at second order, the
calculation of the probability density function reduces to the
calculation of a single quadrature. Moreover, the case where the
volume effects are taken into account only requires the calculation of
an additional integral. The method is then applied to various concrete
cases, as the chaotic, new, hybrid and running mass inflationary
models. In each case, the probability density function can be computed
analytically and the volume effects evaluated exactly. This allows us
to study the relevance of the quantum effects in those models. In
particular, in the case of the running mass model, this is the first
time that such an investigation is carried out.

\par

Our paper is organized as follows. In the next section
(Sec.~\ref{Basic equations}), we briefly review the basic equations of
stochastic inflation. Then, in Sec.~\ref{Solving stochastic
inflation}, we present our method and compare it with the approaches
known in the existing literature. We also show how to compute the
probability density function, with or without the volume effects taken
into account, directly from the Langevin equation without writing a
Fokker-Planck equation. As already mentioned, we explicitly
demonstrate that this calculation simply reduces to the calculation of
a single quadrature. In Sec.~\ref{Inflationary models}, we briefly
present the inflationary models the stochastic effects of which are
computed in the subsequent section. In particular, we focus on the
choice of the free parameters characterizing the corresponding
potentials since their values are crucial in order to estimate the
importance of the quantum effects. In Sec.~\ref{Results and
Discussion}, we discuss and interpret our results for the cases of
large field models, small field models, hybrid inflation and running
mass inflation. To our knowledge, the calculation of the mean value
and the variance of the coarse-grained field in the last three models
was never done before (when the backreaction is taken into
account). We end this article by some concluding remarks.


\section{Basic equations}
\label{Basic equations}

In the Friedmann-Lema\^{\i}tre-Robertson-Walker (FLRW) Universe, the
assumptions of homogeneity and isotropy allow us to write the metric
in the simple form ${\rm d}s^2=-{\rm d}t^2+a^2(t) \delta _{ij}{\rm
d}x^{i}{\rm d}x^j$, where $a(t)$ is the time-dependent scale factor
and where we have assumed flat space-like sections. In such a
space-time the evolution of an homogeneous scalar field $\phi (t)$,
sourcing the metric evolution, is described by the Klein-Gordon
equation
\begin{equation}
\label{KG}
  \ddot\phi+3H\dot\phi+V'(\phi)=0\, ,
\end{equation}
where a dot means a derivation with respect to the cosmic time $t$ and
a prime a derivation with respect to the scalar field $\phi $. This
equation is coupled to the Friedmann equation for the scale factor
\begin{equation}
  H^2\equiv\left(\frac{\dot a}{a}\right)^2 =\frac{\kappa
  }{3}\left[\frac{\dot\phi^2}{2}+V(\phi)\right]\, ,
\end{equation}
where we have defined $\kappa\equiv 8\pi/\mpl^2$, $\mpl$ being the
Planck mass.

\par

During a phase of inflation, if the slow roll approximation is
satisfied then the acceleration $\ddot\phi$ of the field is negligible
compared to the friction term $3H\dot\phi$ and, at the same time, the
kinetic energy $\dot\phi^2/2$ is small compared to the potential
energy $V(\phi)$. This approximation considerably simplifies the
equations describing the evolution of the system. These ones can now
be re-written as
\begin{equation}
\label{slowroll}
H^2(\phi)=\frac{\kappa}{3}V(\phi)\, , \quad
\dot\phi+\frac{2}{\kappa}H'\!(\phi)=0\, .
\end{equation}
At the classical level, there is nothing more to say. Once we are
given a potential $V(\phi)$, the above equations can be solved and the
time evolution of the scale factor and of the inflaton field obtained.

\par

The problem becomes much more complicated when the field $\phi $ is
considered as a quantum operator. A first difficulty arises because
quantizing a scalar in curved space-time is technically complicated
for the case of an arbitrary potential (for an arbitrary potential,
the Klein-Gordon equation is non-linear and one cannot Fourier expand
the field and write an equation for a time-dependent mode function). A
second (more fundamental) difficulty is to take into account the
effect of the quantum scalar field on the geometry, \ie the
backreaction problem. Since the inflaton sources the Einstein
equations, the Friedmann equation (taken literally) indicates that the
geometry should also be quantized. Unfortunately, this quantum gravity
regime is presently not known and the previous program cannot be
carried out.

\par

The stochastic formalism allows us to circumvent these
difficulties. In the stochastic formalism~\cite{stocha}, one is
interested in the dynamics of a ``coarse-grained'' field $\varphi
(t,{\bf x})$. This coarse-grained field is defined to be the spatial
average of the ordinary field $\phi$ over a physical volume the size
of which is typically larger than the Hubble radius $H^{-1}\equiv
a/\dot{a}$. Therefore, $\varphi(t,{\bf x})$ basically contains the
long-wavelength Fourier modes (\ie those with comoving wavenumber such
that $k<aH$) only.

\par

The evolution of the coarse-grained field is still described by the
Klein-Gordon equation~\eqref{KG} but a suitable random noise field
$\xi (t)$, acting as a classical stochastic source term, should be
added in order to take into account the effect of the quantum
fluctuations. In the slow-roll approximation, the evolution of the
coarse-grained field is thus governed by a first order Langevin-like
differential equation which can be written as~\cite{early, stocha,
rey,NNS,Kandrup,NS,N}
\begin{equation}
\label{Langevin}
  \dot\varphi +\frac{1}{3H}\frac{{\rm d}V}{{\rm
  d}\varphi}= \frac{H^{3/2}}{2\pi}\xi(t)\, ,
\end{equation}
where the noise field $\xi$ is defined in such a way that its mean and
two-point correlation function simply read
\begin{equation}
\label{noisecorr}
  \mean{\xi(t)}=0,\quad
  \left\langle\xi(t)\xi(t')\right\rangle=\delta(t-t')\, ,
\end{equation}
$\delta(z)$ being the Dirac distribution. The normalization of the
correlation function is chosen in order to reproduce the ordinary
result, $\mean{\varphi^2}=H^3t/(4\pi^2)$, valid for a free field in de
Sitter space-time.

\par

It is at this point that the backreaction problem shows up. The
standard assumption is that the Hubble parameter in
Eq.~(\ref{Langevin}) is only controlled by the coarse-grained
field. Then, one needs to specify how $H$ depends on $\varphi$ and one
naturally assumes that the Friedmann equation (in the slow-roll
approximation) holds for the coarse grained quantities, namely
\begin{equation}
\label{einstein}
  H^2(\varphi)\simeq \frac{\kappa}{3}V(\varphi)\, .
\end{equation}
A direct consequence of the above equation is that the noise becomes
multiplicative. The formalism briefly described previously also
indicates that the coarse-grained field $\varphi$ now describes a
Brownian motion for which the classical drift is modified by the
quantum diffusion term. The main goal is to solve Eq.~(\ref{Langevin})
since, endowed with the solution, we can then evaluate the probability
density function of the coarse-grained field and/or various
correlation functions.


\section{Solving stochastic inflation}
\label{Solving stochastic inflation}

\subsection{Perturbative Method}

As discussed above, the main purpose of this article is to present and
study a method for solving the Langevin equation. This method was used
for the first time in Ref.~\cite{GLMM} for the calculation of CMB
non-Gaussianities and in Ref.~\cite{MM1} in order to compute how the
quantum effects affect the behavior of the quintessence field during
inflation. Here, we develop the method in full generality including
the calculation of the probability density function. The main idea is
to consider the coarse-grained field $\varphi$ as a perturbation of
the classical solution $\varphi _{\rm cl}$ (the classical solution is
defined as the solution of the Langevin equation without the noise),
$\varphi_\cl$ being supposed to be known. The corrections to $\varphi
_{\rm cl}$ are obtained by adding successive terms of higher and
higher powers in the noise, \ie
\begin{equation}
  \varphi(t) = \varphi_\cl(t)+\df_1(t)+\df_2(t)+\cdots \, ,
\end{equation}
where the term $\df_i(t)$ depends on the noise at the power $i$. The
equations of motion controlling the evolution of the term $\df_i(t)$
are obtained by inserting the above expansion into the Langevin
equation and by identifying the terms of same order. Expanding up to
second order, we get two linear differential equations for $\df_1$ and
$\df_2$, namely
\begin{equation}
\label{eqdefi1}
  \frac{{\rm d}\df_1}{{\rm d}t} + \frac{2}{\kappa}H''(\varphi_{\rm
  cl}) \df_1 = \frac{H^{3/2}(\varphi _{\rm cl})}{2\pi}\xi(t)\, ,
\end{equation}
and
\begin{align}
\label{eqdefi2}
  \frac{{\rm d}\df_2}{{\rm d}t} &+ \frac{2}{\kappa}H''(\varphi
  _{\rm cl}) \df_2 = -\frac{H'''(\varphi _{\rm cl})}{\kappa}\df_1^2
  \nonumber \\ &+ \frac{3}{4\pi}H^{1/2}(\varphi _{\rm
  cl})H'(\varphi _{\rm cl})\df_1\xi(t)\, .
\end{align}
These equations are similar to Eq.~(17) of Ref.~\cite{GLMM}. The only
difference is that, in the previous reference, the Langevin equation
is written in term of the number of e-folds while, here, our time
variable is the cosmic time. Since the above equations are linear,
they can be solved by varying the integration
constant. Straightforward manipulations lead to
\begin{equation}
\label{slowfirst}
  \df_1(t)=\frac{H'\left[\varphi _{\rm
  cl}(t)\right]}{2\pi}\int_\tin^t\!\!\dd\tau
  \frac{H^{3/2}\left[\varphi _{\rm cl}(\tau )\right]}{H'\left[\varphi
  _{\rm cl}(\tau )\right]}\xi(\tau)\, ,
\end{equation}
where we have assumed that the initial conditions are such that
$\df_1(t=\tin)=0$. In the same manner, the solution for $\df_2(t)$ can
be easily obtained and reads
\begin{align}
\label{slowsecond}
  \df_2(t) = & -\frac{H'}{\kappa }\int _\tin^t \!\!\dd\tau
  \frac{H'''}{H'}\delta \varphi _1^2(\tau ) \notag \\ 
  & +\frac{3H'}{4\pi}\int_\tin^t\!\!\dd\tau
  H^{1/2}\df_1(\tau)\xi(\tau) \, .
\end{align}
As expected, $\df_1$ is linear in the noise $\xi$ while $\df_2$ is
quadratic. Of course, the expansion could be pushed further and one
could evaluate $\df_3$, $\df_4 $ etc \dots using the same technique.

\par

We are now in a position where the various correlation functions can
be calculated exactly. Since $\df_1$ is linear in the noise, its mean
value obviously vanishes
\begin{equation}
  \mean{\df_1}=0\, .
\end{equation}
This means that $\df_1$ does not introduce any correction to the mean
value of the coarse-grained field. As a matter of fact, $\df_1$
directly contributes only to the variance of $\varphi $. Using the
white noise correlation function given by Eq.~(\ref{noisecorr}), we
obtain
\begin{equation}
\label{slowvar}
  \mean{\df_1^2} = 
  \frac{\kappa}{2}\left(\frac{H'}{2\pi}\right)^2 
  \!\!\int^{\varphi_\ini}_{\varphi _{\rm cl}}\!\!\dd\psi
  \left(\frac{H}{H'}\right)^3\, .
\end{equation}
We see that the calculation of the variance reduces to a simple
quadrature. In order to calculate the correction to the mean value we
must consider the mean of $\df_2$. Using the fact that
$\mean{\df_1(\t)\xi(\t)}=H^{3/2}/(4\pi)$, we arrive at
\begin{align}
  \mean{\df_2} = & \,\frac{H'}{2\pi\mpl^2}\Bigg\{H''\!
  \int^{\varphi_\ini}_{\varphi_\cl}\!\!\!\dd\psi\left(\frac{H}{H'}\right)^3\notag\\
  &- \!\int^{\varphi_\ini}_{\varphi _\cl}\!\!\!\dd\psi
  \bigg[H''\left(\frac{H}{H'}\right)^3 -
  \frac{3}{2}\frac{H^2}{H'}\bigg]\Bigg\}\, .
\end{align}
In this expression, the first term is nothing but the one given in
\eqref{slowvar}, \ie $\mean{\df_1^2}$, while the second one can be
evaluated exactly with the help of an integration by parts. This leads
to 
\begin{equation}
\label{slowmean}
  \mean{\df_2} = \frac{H''}{2H'}\mean{\df_1^2}
  +\frac{H'}{4\pi\mpl^2}\Bigg[\frac{H_\ini^3}{(H_\ini')^2}
  -\frac{H^3}{(H')^2}\Bigg]\, .
\end{equation}
Therefore, at second order in the noise, everything can be reduced to
the calculation of a single quadrature, the one of
Eq.~(\ref{slowvar}). Before discussing the probability density
function, we compare the method described above with what is already
known in the literature.

\subsection{Comparison with Other Methods}

As already mentioned before, in order to treat the case with
backreaction, various papers~\cite{Hodges,YVM} concentrate on very
particular cases where the Langevin equation can be solved exactly
(see also the recent article, Ref.~\cite{GT}, where the same method is
used). The typical example of this procedure is the model described by
a quartic potential, $V=3\lambda _4\varphi ^4/(8\pi)$, $\lambda _4 $
being a dimensionless coupling constant, for which the Langevin
equation can be written as
\begin{eqnarray}
\label{Langevin2}
\frac{{\rm d}}{{\rm d}t}\left(\frac{\varphi
}{\mpl}\right)+\frac{\sqrt{\lambda _4}}{2\pi }\varphi = \frac{\lambda
_4^{3/4}\mpl^{1/2}}{2\pi}\left(\frac{\varphi
}{\mpl}\right)^{3}\xi(t)\, .
\end{eqnarray} 
This equation can be solved exactly because it takes the form of a
Bernoulli equation after a change of variable. However, in this case,
one does not obtain the coarse-grained field itself but rather some
power of it, namely
\begin{eqnarray}
\label{solphi}
  \varphi^{-2}(t) = \varphi_\cl^{-2}(t) \left[1 - \Psi(t)\right]\, ,
\end{eqnarray}
where $\varphi_\cl(t)$ is the classical solution (which, in the
slow-roll approximation, is known explicitly) and where the stochastic
quantity $\Psi (t)$ is defined by
\begin{equation}
  \Psi (t)\equiv \frac{\lambda _4^{3/4}\mpl^{1/2}}{\pi}
  \int_{t_\ini}^t{\rm d}\tau \left[\frac{\varphi
  _\cl(\tau)}{\mpl}\right]^2\xi(\tau )\, , 
\end{equation}
which is a new dimensionless Gaussian noise with vanishing mean value
and whose variance (and higher correlation functions) can easily be
computed. Therefore, if one wants to obtain the field itself, it is
necessary to take the inverse square root of the
solution~(\ref{solphi}).

\par

At this point, several remarks are in order. First, the only way to
compute the coarse-grained field $\varphi $ and its various
correlation functions is to expand $(1-\Psi)^{-1/2}$ in $\Psi $, that
is to say in the coupling constant $\lambda _4$, and to truncate the
expansion at some order (the series does not converge anyway). Thus,
we see that, despite the fact that we have an exact solution, an
expansion is still required in order to use Eq.~(\ref{solphi})
concretely. Second, one can show that the expansion in the coupling
constant is equivalent to our expansion in the
noise~\cite{MM1}. However, clearly, our method is more general because
it is not restricted to the situation where an exact solution of the
Langevin equation is available. This is because the expansion is
directly performed in the Langevin equation rather than in its
solution. The drawback of the method used in Refs.~\cite{Hodges,YVM}
is clearly that it is first necessary to find a solution of the
Langevin equation before the expansion can be taken. We will
illustrate this last remark on the calculation of the criterion which
determines when the self-reproducing regime becomes efficient (\ie
when the quantum fluctuations dominate the classical drift). In
Ref.~\cite{GT}, using the model $V\propto \varphi ^4$ for the reasons
described before, the authors have recovered the standard result that
this happens when the initial value of the inflaton is larger than
$\varphi _\ini \sim \lambda _4^{-1/6} \mpl$. Using our formalism, we
will derive this criterion for any model of the form $V\propto \varphi
^n$, which would not have been possible with the method of
Ref.~\cite{GT}.

\par

Another possibility studied in the literature is the so-called scaling
solutions method, see Ref.~\cite{scaling}. The idea is to perform a
change of variable and to work in terms of a new stochastic process
$\eta (\varphi )$ such that the Langevin equation takes the form ${\rm
d}\eta/{\rm d}t ={\cal F}(t;\eta )\xi (t)$, where ${\cal F}$ is {\it a
priori} a complicated function of $\eta $. If $\eta $ is replaced by
$\eta _\cl$ in ${\cal F}$, then the new Langevin equation becomes
solvable. Therefore, one sees that this method bears some resemblance
with the method investigated here. However, there also exists
important differences. First, our method does not require any change
of variable which is an advantage since, in general, the link between
$\varphi $ and $\eta $ is quite complicated. Second, in the scaling
method, there is no systematic expansion in the noise (in some sense
one always works at first order) while in our method we can go to any
order, the only limitation being mathematical complexity. Third, a
saddle point approximation is used to estimate the effective
dispersion while, in our case, following the calculations of the
previous subsection, this can be done exactly. Fourth, it is difficult
to evaluate the reliability of the scaling limit while we will discuss
in a forthcoming article~\cite{MM3} how the accuracy of our method can
be determined precisely. Finally, let us also stress that, in
Ref.~\cite{scaling}, only the cases of large field models and
exponential potentials are considered (in principle, it would be
possible to treat other models with the scaling method although it is
unclear whether this would lead to analytical expressions for the
probability density) while, here, we will also apply our method to the
new, hybrid and running mass inflationary scenarios.

\par

Finally, let us repeat that the method studied here is similar to the
one used in Ref.~\cite{GLMM} even if we apply it in a different
context. In particular, the solution given by Eqs.~(\ref{slowfirst})
and (\ref{slowsecond}) are identical to Eqs.~(19) and~(21) of
Ref.~\cite{GLMM}, these formulas, however, being written in terms of
the total number of e-folds rather than in terms of cosmic time. In
Ref.~\cite{GLMM}, these results are applied to the calculation of the
three-point correlation function of the CMB fluctuations while, in the
present article, we use them, among others, in order to derive the
probability density function of the background field.

\par

To end this subsection, let us recall that more details on the other
methods discussed here can be found in Ref.~\cite{MM1}.

\subsection{Probability Density Function}

After having compared our approach with other formalisms, we now come
back to the general method and show how one can calculate the single
point probability distribution $P_{\rm c}(\varphi,t)$ of the
coarse-grained field (also sometimes called $P_{\rm p}$ in the
literature). Let us recall that $P_{\rm c}(\varphi,t)$ is the
probability of the stochastic process to assume a given value at a
given time in a single coarse-grained domain. Very often, this
probability distribution is obtained from a Fokker-Planck
equation. However, it can also be determined from~\cite{pdf}
\begin{equation}
\label{defpdf}
P_{\rm c}(\varphi,t) = \mean{\delta(\varphi-\varphi[\xi])}\, .
\end{equation}
where $\varphi[\xi]$ is the solution of the Langevin equation, and the
mean value has to be evaluated with the functional probability
distribution $\mathcal{P}[\xi]$ of the noise 
\begin{equation}
  \mathcal{P}[\xi]=\mathcal{N}_0
  \exp\left[-\frac{1}{2}\xi^{\rm T} \mathbf{C}^{-1}\xi\right],
\end{equation}
with the normalization $\mathcal{N}_0$ simply being given by
$\mathcal{N}_0=(\int\!\mathcal{D}\xi\,\e^{-\xi^{\rm T}
\mathbf{C}^{-1}\xi/2})^{-1}$ and where we have introduced the
definition $f^{\rm T}g\equiv\int \dd\t f(\t)g(\t)$. As one can see on
the above expression, the noise probability distribution is Gaussian
and correctly yields the noise correlation function
$\mean{\xi(t)\xi(t')}=\mathbf{C}(t,t')$. In Eq.~(\ref{defpdf}), the
stochastic field will be given by our perturbative solution, namely
$\varphi[\xi]=\varphi _\cl +\df_1+\df_2$. The first and second order
corrections to the classical solution, which are linear and bilinear
in the noise respectively, can also be written as $\df_1 = J^{\rm
T}\xi $ and $\df_2 = \xi^{\rm T}\mathbf{A}\xi$, where the detailed
definition of $J$ and $\mathbf{A}$ are given in
Appendix~\ref{appvolume}. Then, using the integral representation of
the $\delta$ function, we get
\begin{widetext}
\begin{equation}
  P_{\rm c}(\varphi,t) = \frac{1}{2\pi }\int \dd
  y\exp\left[iy\left(\varphi_\cl-\varphi\right)\right] {\cal
  N}_0\int\mathcal{D} \xi\,\exp\left[ -\frac{1}{2}\xi^{\rm T}
  (\mathbf{C}^{-1}-2iy\mathbf{A})\xi + iy J^{\rm T}\xi \right]\, .
\end{equation}
In this expression the functional integration is a Gaussian
integration involving the kernel
$\mathbf{C}^{-1}-2iy\mathbf{A}$. Defining the new coefficient
$\mathcal{N}_y=[\int\!\mathcal{D}\xi\,\e^{-\xi^{\rm T}
(\mathbf{C}^{-1}-2iy\mathbf{A})\xi/2}]^{-1}$, the functional
integration yields
\begin{equation}
\label{funcint}
 P_{\rm c}(\varphi,t) = \frac{1}{2\pi }\int \dd y\exp
  \left[iy\left(\varphi_\cl-\varphi\right)\right] \frac{{\cal
  N}_0}{{\cal N}_y}\exp\left[ -\frac{1}{2} y^2 J^{\rm T}
  (\mathbf{C}^{-1}-2iy\mathbf{A})^{-1}J \right]\, .
\end{equation}
\end{widetext}
Then, generalizing the relation valid for two finite $n\times n$
matrices $\mathbf{M}$ and $\mathbf{N}$,
\begin{equation}
  \frac{\displaystyle\int \dd^n x \,\e^{-x_i M_{ij} x_j}}
  {\displaystyle\int \dd^n x \,\e^{-x_i N_{ij} x_j}}
  = \sqrt{\frac{\det \mathbf{N}}{\det \mathbf{M}}}
  = \e^{-\tr\ln \mathbf{N^{-1}}\mathbf{M}/2}\, ,
\end{equation}
to the continuous case~\cite{zinn}, we can write the ratio of the two
normalization coefficients as
\begin{align}
  \frac{\mathcal{N}_0}{\mathcal{N}_y}
  &=
  \exp\left[-\frac{1}{2}\tr\ln(\mathbf{1}-2iy\mathbf{AC})\right]\simeq
  \e^{iy\mean{\df_2}}\, ,
\end{align}
up to second order in the noise. Finally, if the argument of the
second exponential in Eq.~(\ref{funcint}) is expanded in powers of the
noise, then all the terms but $J^{\rm T} \mathbf{C}J=\mean{\df_1^2}$
can be neglected.  Evaluating the remaining ordinary integration over
$y$ we get the normalized Gaussian distribution
\begin{equation}
\label{pc}
  P_{\rm c}(\varphi,t) = \frac{1}{\sqrt{2\pi\!\mean{\df_1^2}}}
  \exp\!\left[
  -\frac{(\varphi-\varphi_\cl-\mean{\df_2})^2}{2\mean{\df_1^2}}\right]\,
  .
\end{equation}
This distribution is centered over the mean value
$\mean{\varphi}\simeq\varphi_\cl + \mean{\df_2}$ with variance
$\mean{\df_1^2}$. The above equation is one of the main result of this
article. In order to evaluate $P_{\rm c}$ only the
integration~(\ref{slowvar}) is necessary which illustrates the power
of the perturbative method.

\subsection{Volume Effects}

If we want to investigate the evolution of the probability
distribution of the field when spatially averaged over the entire
Universe (and not only in the single domain) we must take into account
the volume effects.  These are due to the fact that the size of each
homogeneous domain depends on the value of the field within the domain
itself, and we expect larger domains to give a more important
contribution to the average over space. The value of each
field-dependent quantity must thus be weighted with the physical
volume $a^3(\varphi)=\exp[3\int {\rm d}\tau H(\varphi )]$.  Therefore,
the normalized probability distribution accounting for volume effects
can be obtained from
\begin{equation}
\label{volume}
  P_{\rm v}(\varphi,t) =
  \frac{\mean{\delta(\varphi-\varphi[\xi])\,\e^{3\!\int\dd\t
  H(\varphi[\xi])}}} {\mean{\e^{3\!\int\dd\t H(\varphi[\xi])}}}\, .
\end{equation}
In order to compute this new distribution function, we expand
perturbatively $H(\varphi[\xi])$ up to second order in the noise,
\begin{equation}
H(\varphi[\xi]) = H_\cl+H'_\cl(\df_1+\df_2)
+\frac{H''_\cl}{2}\df_1^2\, ,
\end{equation}
and write the first order term appearing in the argument of the
exponential in Eq.~(\ref{volume}) as $ \int_{t_\ini}^t\dd\t
H'(\t)\df_1(\t) \equiv I^{\rm T}\xi$ while the second order ones take
the form $ \int_{t_\ini}^t\dd\t \left[H'(\t)\df_2(\t) +
H''(\t)\df_1^2/2\right] \equiv \xi^{\rm T} \mathbf{B} \xi $. 

\par

Then, one can repeat the calculations performed in the previous
subsection and we obtain the following expression for the volume
weighted distribution function
\begin{widetext}
\begin{equation}
  P_{\rm v}(\varphi,t) = \frac{1}{2\pi \mean{a^3(\varphi)}}\int \dd
  y\exp\left[iy\left(\varphi_\cl-\varphi\right)\right] {\cal
  N}_0\int\mathcal{D} \xi\,\exp\left[ -\frac{1}{2}\xi^{\rm T}
  (\mathbf{C}^{-1}-2iy\mathbf{A}-6\mathbf{B})\xi + \left(iy J^{\rm
  T}+3I^{\rm T}\right) \xi \right]\, .
\end{equation}
\end{widetext}
This is again a Gaussian integration but with the modified kernel
$\mathbf{C}^{-1}-2iy\mathbf{A}-6\mathbf{B}$ where the term
$-6\mathbf{B}$ is a new contribution that accounts for the volume
effects. These volume effects also manifest themselves in the term
linear in the noise (\ie the term proportional to $I$). Following the
same steps as before, we define the new normalization $\mathcal{N}'_y
= \left[\int\!\mathcal{D}\xi\,\e^{-\xi^T
(\mathbf{C}^{-1}-2iy\mathbf{A}-6\mathbf{B})\xi/2}\right]^{-1}$ and
evaluate the functional integral. We obtain
\begin{widetext}
\begin{equation}
\label{inter}
P_{\rm v}(\varphi,t) = \frac{1}{2\pi \mean{a^3(\varphi )}}\int \dd
  y\exp \left[iy\left(\varphi_\cl-\varphi\right)\right] \frac{{\cal
  N}_0}{{\cal N}'_y}\exp\left[ \frac{1}{2} (iyJ^{\rm T}+3I^{\rm T})
  \left(\mathbf{C}^{-1}-2iy\mathbf{A}-6\mathbf{B}\right) ^{-1}(iyJ+3I)
  \right]\, .
\end{equation}
\end{widetext}
As in the previous case, the functional inverse of the modified kernel
$(\mathbf{C}^{-1}-2iy\mathbf{A}-6\mathbf{B})^{-1}$ simply reduces, up
to second order, to $\mathbf{C}$.

\par

We must now evaluate the denominator. Exactly in the same way as
before, this term becomes
\begin{equation}
  \mean{a^3(\varphi)\!}={\cal N}_0\!\!\int\!\mathcal{D} \xi\exp\!\left[
  -\frac{1}{2}\xi^{\rm T}\! (\mathbf{C}^{-1}\!\!\!-6\mathbf{B})\xi + 3I^{\rm
  T}\!\xi\right]\! ,
\end{equation}
and the functional integration yields $\mean{a^3(\varphi)} \simeq
(\mathcal{N}_0/\mathcal{N}'_0)\exp\left(9I^{\rm T} \mathbf{C} I/2
\right)$, where $\mathcal{N}'_0$ is a new normalization coefficient
the explicit expression of which we do not give here for
simplicity. Then, we insert this last expression in
Eq.~(\ref{inter}). As expected the terms ${\cal N}_0$ cancels
out. Moreover, when evaluating the ratio
$\mathcal{N}'_0/\mathcal{N}'_y$, the contributions of the terms
describing the volume effects (those involving $\mathbf{B}$) also
cancel out at second order and we get $\mathcal{N}'_0/\mathcal{N}'_y
\simeq\e^{iy\mean{\df_2}}$ as before. Putting everything together, we
obtain the final result
\begin{equation}
\label{pv}
  P_{\rm v}(\varphi,t) = \frac{1}{\sqrt{2\pi\!\mean{\df_1^2}}}
  \exp\!\left[ -\frac{\left(\varphi-\mean{\varphi}-\!3I^{\rm
  T}\!\!J\right)^2}{ 2\mean{\df_1^2}}\right] ,
\end{equation}\\[0pt]
where $\mean{\varphi}\!=\varphi_\cl+\mean{\df_2}$ is the usual mean value.

\par

This expression should be compared with Eq.~(\ref{pc}). The variance
of the resulting Gaussian probability distribution is thus unchanged,
while the volume-weighted mean value $\mean{\varphi}_{\rm v}=\mean{\varphi}
+ 3\,I^{\rm T}\!J$ gets the extra correction 
\begin{equation}
\label{volcorrection}
  3\,I^{\rm T}\!J=3\!\int_{t_\ini}^t\!\dd\t
  H'(\t)\mean{\df_1(t)\df_1(\t)}\, .
\end{equation}

So far, the calculation of the volume effects, in particular
Eqs.~(\ref{pv}) and (\ref{volcorrection}), do not rely on the
slow-roll approximation. However, if this approximation is satisfied,
then the volume contribution can be easily calculated for a generic
potential. We obtain
\begin{equation}
\label{voleffects}
  3I^{\rm T}\!J= \frac{12H'}{\mpl^4}
  \!\int_{\varphi_\cl(t)}^{\varphi_\ini}\!\!\dd\psi\frac{H^4}{(H')^3}
  -12\pi\frac{H}{H'}\frac{\mean{\df_1^2(t)}}{\mpl^2}\, .
\end{equation}
Therefore, we see that the calculation of the volume effects only
requires the computation of one additional quadrature.


\section{Inflationary models}
\label{Inflationary models}

In this section, we briefly present the inflationary models to which
our method is applied in the next section. In particular, we carefully
discuss the choice of the free parameters characterizing those models
since their numerical values turn out to be crucial in order to
estimate the importance of the stochastic effects.

\subsection{Generalities}

We adopt a parameterization of $V(\phi)$ suitable for describing
different types of inflationary models, and we write the potential as
\begin{equation}
\label{potential}
  V(\varphi) = M^4\!\left[a + b
  \left(\frac{\varphi}{\mu}\right)^{\!\!n}\right]\, ,
\end{equation}
where $a=0,1$ and $b =\pm 1$ according to the case under
consideration, while $M$, $\mu $ and $n$ (with $n\geq 2$) are free
parameters. If $a=0$ and $b=1$ we have monomial potentials describing
chaotic inflation~\cite{chaotic}, also commonly known as ``large field
models'' (LF) because the initial value of the field (rolling towards
the origin) is typically much larger than $\mpl$. The case $n=4$ has
already been treated in Refs.~\cite{Hodges,YVM} while the general
case, \ie for an arbitrary value of $n$, was studied for instance in
Refs.~\cite{early,N,MM1}.  Quantum effects for potentials with $a=1$
have not been computed explicitly before and, therefore, we will
mainly focus on those examples. Potentials with $a=1$ and $b=-1$
belongs to the class of the ``small fields models'' (SF) such as the
new inflation scenario, where the field starts in the false vacuum
close to the origin and moves down to $\phi=\mu$ as in a spontaneous
symmetry breaking~\cite{new}. At this point, one remark is in
order. In fact, the case of stochastic new inflation has been
investigated many times in the literature, for instance in
Ref.~\cite{NNS}. But usually, and this is in this sense that the
treatment presented here is new, the backreaction is not taken into
account and the Hubble parameter is just considered as a constant. In
this article, we do not make this assumption. Finally, the case $a=1$
and $b=1$ describes hybrid inflation~\cite{hybrid}. Although hybrid
inflation is a two-field model, the slow-rolling phase taking place in
the inflationary valley of the potential can effectively be described
as a single field model.

\par

We also consider the running-mass model
(RM)~\cite{running,runningmore}, the potential of which does not
belong to the class presented above. For this model $V$ is given
by~\cite{running}
\begin{equation}
\label{potentialrunning}
  V(\varphi) = M^4\left[1-\frac{c}{2}\left(-\frac{1}{2} +\ln
\frac{\varphi }{\varphi _0}\right)\frac{\varphi ^2}{M_{_{\rm
Pl}}^2}\right],
\end{equation}
where $M_{_{\rm Pl}}\equiv \mpl/\sqrt{8 \pi}$. In this expression, $M$, 
$c $ and $\varphi _0$ are free parameters (In Ref.~\cite{running},
$M^4$ is denoted $V_0$ and $\varphi _0$ is written $\phi _*$). Let us
notice that $c$ can be positive or negative.

\par

Our next step consists in obtaining the classical trajectory for these
models. This can be done if the slow-roll approximation is satisfied
but, even in this case, the classical trajectory can be found
implicitly only. In terms of total number of e-folds $N$, we have for
the models described by Eq.~(\ref{potential})
\begin{equation}
  N=-\kappa \frac{\mu ^2}{n b}
  \int_{\varphi_\ini/\mu}^{\varphi_\cl/\mu} \!\!\dd x\,x^{1-n} \left(a
  + b x^n\right)\, .
\end{equation}
The integration can easily be performed and the solution can be
expressed as (in the following we use the fact that, when
non-vanishing, $a$ is one and that $b$ is just a sign)
\begin{align}
  N &= \,\kappa \frac{\mu ^2}{2n}\,\Bigg\{
  \left(\frac{\varphi_\ini}{\mu}\right)^2 -
  \left(\frac{\varphi_\cl}{\mu }\right)^2 \notag\\ &-
  \frac{2\,ab}{n-2}\left[\left(\frac{\varphi_\ini}{\mu
  }\right)^{\!2-n} - \left(\frac{\varphi _\cl}{\mu
  }\right)^{\!2-n}\right]\Bigg\}\, ,
\end{align}
for $n\neq2$, while for $n=2$ one has
\begin{equation}
  N = \kappa \frac{\mu
  ^2}{4}\left[\left(\frac{\varphi_\ini}{\mu}\right)^2 -
  \left(\frac{\varphi_\cl}{\mu }\right)^2 - ab \ln \left(\frac{\varphi
  _\cl }{\varphi _\ini}\right)^2\right].
\end{equation}
If $a=0$ it is very easy to find the field evolution inverting the
above expressions and solving for $\phi$. This was done, for instance,
in Ref.~\cite{MM1}. On the contrary, if $a=1$ an explicit solution can
be found only for particular values of $n$.  For simplicity, we
concentrate on the specific case $n=2$ for which we get
\begin{equation}
  \frac{\varphi _\cl}{\mu} = \sqrt{b\, W_0\!
  \left\{b\!\left(\!\frac{\varphi_\ini}{\mu}\!\right)^{\!\!2}\!
  \exp\!\left[\frac{\varphi_\ini^2-(N/2\pi)\mpl^2}{b\mu^2}\right]\!
  \right\}}\,,
\end{equation}
where $W_0(x)$ is the principal branch of the Lambert
function~\cite{lambert}.  This special function is the solution of the
equation $W(x)\e ^{W(x)}=x$. Since the curve $x\,\e^x$ has a global
minimum for $x=-1$, its inverse is a multivalued function with two
branches on the real axe (and infinite branches on the complex plane).
The one being continuous through the origin and defined on the real
interval $[-1/\e,\infty)$ is called the principal branch and is
denoted $W_0$. The secondary branch, conventionally chosen to be the
one defined on $[-1/\e,0]$ and denoted $W_{-1}$, diverges at the
origin (such that $W_{-1} \rightarrow -\infty$).  In our case, we have
to choose the principal branch since for $b=1$ the argument of $W$ is
positive, and for $b=-1$ we must have $\phi/\mu<1$.

\par

In the case of the running-mass model~(\ref{potentialrunning}), the
total number of e-folds can also be obtained explicitly. It reads
\begin{widetext}
\begin{align}
N &= \frac{1}{c}\left(\ln \left\vert\ln \frac{\varphi _\cl}{\varphi
    _0}\right\vert-\ln \left\vert\ln \frac{\varphi _\ini}{\varphi
    _0}\right\vert\right) +\frac{1}{4}\left(\frac{\varphi _0}{M_{_{\rm
    Pl}}}\right)^{\!2}\left[ {\rm Ei}\left(2\ln \frac{\varphi
    _\cl}{\varphi _0}\right) -{\rm Ei}\left(2\ln \frac{\varphi
    _\ini}{\varphi _0}\right)\right]
    -\frac{1}{4}\left[\left(\frac{\varphi _\cl}{M_{_{\rm
    Pl}}}\right)^{\!2} -\left(\frac{\varphi _\ini}{M_{_{\rm
    Pl}}}\right)^{\!2}\right] ,
\end{align}
\end{widetext}
where the exponential integral function is defined by~\cite{Grad}
${\rm Ei}(x)\equiv -\int _{-x}^{+\infty} {\rm d}t\e ^{-t}/t$. 
Obviously, this expression is too complicated to be
inverted. However, if, as done in Ref.~\cite{running}, one notices
that $\varphi _\cl /M_{_{\rm Pl}}\ll 1$ then one can just replace
$V$ in the expression giving the number of e-folds by $M^4$. This
leads to
\begin{equation}
\label{Napproxrun}
N\simeq \frac{1}{c}\left(\ln \left\vert \ln \frac{\varphi
    _\cl}{\varphi _0}\right\vert -\ln \left\vert \ln \frac{\varphi
    _\ini}{\varphi _0}\right\vert \right)\, ,
\end{equation}
and then, since the previous expression can be inverted, one obtains
the classical field as a function of the number of e-folds
explicitly, namely
\begin{equation}
\varphi _\cl \left(N\right) =\varphi _0\exp\left({\rm e}^{cN}\ln
\frac{\varphi _\ini}{\varphi _0}\right)\, .
\end{equation}
 Let us notice that the expression~(\ref{Napproxrun}) is in agreement
with, for instance, Eq.~(21) of Ref.~\cite{running}.

\par

Our next move is to find the numerical value of the parameters $M$ and
$\mu$ or $c$ and $\varphi _0$. This can be done from the measurement
of the CMB anisotropy made by the Wilkinson Microwave Anisotropy Probe
(WMAP) satellite, namely from the formula 
\begin{equation}
\label{multi}
  \frac{Q_\mathrm{rms-PS}^2}{T^2} \equiv \frac{5C_2}{4\pi}
  =\frac{1}{60\pi\epsilon_*}\frac{H_*^2}{\mpl^2}
  =\frac{2}{45\epsilon_*}\frac{V_*}{\mpl^4}\, ,
\end{equation}
where $Q_\mathrm{rms-PS}/T$ has been measured to be
$Q_\mathrm{rms-PS}/T\sim 6\times 10^{-6}$. We now discuss the four
cases separately.

\subsection{Large field models}

For $a=0$ we can eliminate the free mass parameter $\mu$ by simply
rescaling the other parameter $M$. We thus obtain a monomial potential
given by
\begin{equation}
  V(\varphi) = M^4\!\left(\frac{\varphi}{\mpl}\right)^n \, .
\end{equation}
In this case, the slow-roll equation of motion leads to a solution
which is completely explicit and reads
\begin{equation}
\label{solLarge}
  \frac{\varphi}{\mpl} =
  \sqrt{\left(\frac{\varphi_\ini}{\mpl}\right)^2 -\frac{n}{4\pi}N}\, .
\end{equation}
The total number of e-folds during inflation is simply given by
$N_{_{\rm T}}=4\pi (\varphi_\ini/\mpl)^2/n-n/4$ and can be very large
if the initial energy density of the inflaton field is close to the
Planck scale $\mpl^4$. The model remains under control only if the
initial energy density is smaller than $\mpl^4$ and this imposes a
constraint on the initial value of the field, namely $\phiin/\mpl
\lesssim (\mpl/M)^{4/n}$. 

For this kind of potential, the slow roll parameter $\epsilon \simeq
\mpl^2/(16\pi )(V'/V)^2$ becomes $\epsilon =
n^2/(16\pi)(\mpl/\varphi)^2$ and inflation stops when $\epsilon=1$,
\ie when the slow-roll conditions are violated. The corresponding
value of the field is $\varphi_\mathrm{end}= n/(4\sqrt{\pi})\mpl$. The
classical solution \eqref{solLarge} allows us to calculate the field
value $\varphi _*$ at Hubble crossing during inflation in terms of
$N_*$, the number of e-folds between the Hubble radius crossing and
the end of inflation. We get $(\varphi_*/\mpl)^2=n(n+4N_*)/(16\pi )$
from which we deduce the corresponding value of the slow-roll
parameter of $ \epsilon_* = n/(4N_*+n)$. Finally, from the WMAP
normalization, see Eq.~(\ref{multi}), we deduce the mass scale $M$
\begin{equation}
  \left(\frac{M}{\mpl}\right)^{\!4} \!=
  \frac{(45/2)n}{(4N_*+n)^{n/2+1}}\!\left(\frac{16\pi}{n}\right)^{\!\!n/2}
  \frac{Q_{\rm rms-PS}^2}{T^2}\, .
\end{equation}
This is the value of $M$ that we use for the calculation of the
quantum effects. From an observational point of view, all the models
such that $n>5$ are now excluded by the WMAP data, the quartic case
being on the border line~\cite{LL}.

\subsection{Small field models}

\begin{figure}
  \includegraphics[width=.95\columnwidth]{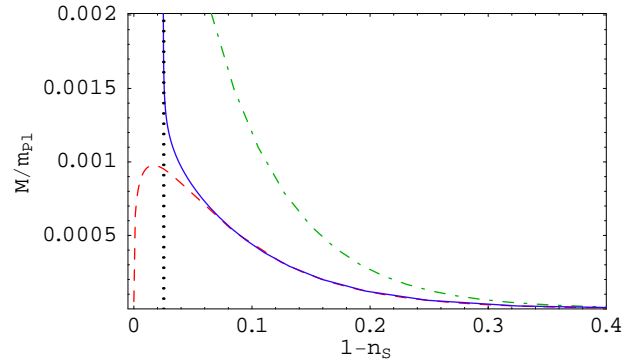}
\caption{\label{M} Characteristic mass scale $M$ obtained from the CMB
normalization as a function of the spectral index for the new
inflationary scenario.  The exact, numerically calculated, curve
(solid blue line) blows up at $1-n_{_{\rm S}}=6/(1+4N_*)$ while the
approximated one (red dashed line) vanishes at the origin. A different
approximation, usually found in the literature (see for instance
Refs.~\cite{KM,JMbraz}), is also shown for comparison (green
dotted-dashed line). The difference between these two lines is
approximatively a factor of $2$ in $M$ and is significant since this
leads to a factor of $16$ in the normalization factor $M^4$ of the
potential and, thus, in the variance and the mean value of the
fluctuations.}
\end{figure}

In this subsection we discuss the WMAP normalization for the
potential~(\ref{potential}) with $a=1$, $b=-1$ and $n=2$. For such a
model the slow roll parameter $\epsilon$ reads
\begin{equation}
\label{espilon}
  \epsilon=\frac{\mpl^2}{4\pi\mu^2}
  \frac{(\varphi/\mu)^2}{\left[1-(\varphi/\mu)^2\right]^2}\, ,
\end{equation}
and, imposing $\epsilon(\phi_\mathrm{end})=1$ at the end of inflation,
we can solve for $\phi_\mathrm{end}$ and obtain
\begin{equation}
\label{end}
  \left(\frac{\phi_\mathrm{end}}{\mu}\right)^2 = 1 +
  \frac{\mpl^2}{8\pi\mu^2} \left(1 -
  \sqrt{1+\frac{16\pi\mu^2}{\mpl^2}}\right)\, .
\end{equation}
Moreover, at Hubble radius crossing, the value of the field can be
expressed exactly as
\begin{equation}
\label{phistar}
  \left(\!\frac{\varphi _*}{\mu}\!\right)^{\!2} \!= -W_0\!\!
  \left[-\!\left(\!\frac{\varphi_\mathrm{end}}{\mu}\!\right)^{\!\!2}
  \!\e^{-(\varphi_\mathrm{end}/\!\mu)^2-\mpl^2\!N_*\!/\!(2\pi \mu^2)}\!
  \right] .
\end{equation}
At this stage, it is interesting to introduce a second slow-roll
parameter, $\epsilon _2$, defined by $\epsilon _2\equiv {\rm d}\ln
\epsilon /{\rm d}N$~\cite{ST}. Then, the spectral index can be written
as $n_{_{\rm S}}=1-2\epsilon -\epsilon _2$, where the slow-roll
parameters are evaluated at Hubble radius crossing. Working out
explicitly the above formulas, one arrives at
\begin{equation}
\label{nnew}
1-n_{_{\rm S}}=\frac{\mpl^2}{2\pi \mu^2} \frac{1+2\left(\varphi _*/\mu
\right)^2}{\left[1-\left(\varphi _*/\mu \right)^2\right]^2}\, .
\end{equation}
Then, one can take the following route. If one chooses a value for the
scale $\mu $, then one can calculate $\varphi _\mathrm{end}$ with
Eq.~(\ref{end}), $\varphi _*$ with Eq.~(\ref{phistar}) and, finally,
the spectral index with the previous formula. Hence, instead of
working with $\mu $, one can express everything in terms of $n_{_{\rm
S}}$. Finally, using Eq.~(\ref{multi}), one can determine the scale
$M$ in terms of $\varphi _*$ and $\mu $ or, equivalently, in terms of
the spectral index. In other words, we end up with an exact relation
$M(n_{_{\rm S}})$. It is represented in Fig.~\ref{M} (solid blue
line). As one can notice on this figure, the curve blows up at
$n_{_{\rm S}}=1-6/(1+4N_*)$. Let us try to understand this behavior in
more details. If we expand the expression giving the spectral index in
terms of the parameter $\mpl/\mu $, one obtains $n_{_{\rm
S}}=1-6/(1+4N_*)+\mpl /[\mu(1+4N_*)^2\sqrt{\pi }]+{\cal O}\left(\mpl
/\mu \right)$ from which we can obtain an expression of $\mpl/\mu $ in
terms of $n_{_{\rm S}}$. In the same manner, one can expand $M$ and
the result reads
$\left(M/\mpl\right)^4=45\pi^2/[(1+4N_*)^{3/2}(\mpl/\mu)]\left(Q_{\rm
rms-PS}/T\right)^2+{\cal O}[(\mpl/\mu )^{3}]$. Putting these two
formulas together, one finally gets that
\begin{align}
  \left(\frac{M}{\mpl}\right)^4 \simeq
  &\frac{45\pi^{3/2}}{(1+4N_*)^{7/2}} \left(\frac{Q_{\rm
  rms-PS}}{T}\right)^2 \notag \\ &\times \!\left(1-n_{_{\rm
  S}}-\frac{6}{1+4N_*}\right)^{\!-1}
\end{align}
up to terms of order $\mathcal{O}\left[(\mpl/\mu)^{3}\right]$, and
we now understand the presence of the singularity. Although exact for
potentials of the form~(\ref{potential}) (with $a=1$ and $b=-1$) , the
behavior of $M$ is this regime is not realistic for the following
reason: on general grounds, it is clear that it is necessary to
consider additional terms in the potential~(\ref{potential}) because,
otherwise, $\varphi =\mu $ is not a minimum. If, during slow-roll
inflation $\varphi \ll \mu $, then these terms are not important for
the calculation of the perturbations. But, if $\mu \gg \mpl$, then one
has that $\phi _{\rm end}\sim \mu $ and we expect the extra terms to
play a role even during slow-roll inflation. In this regime, the shape
of the potential that we used in order to obtain that $M$ has a
singularity is therefore not realistic. Moreover, it has been shown in
Ref.~\cite{GRS} that the presence of these extra terms can strongly
modify the energy scale of inflation.

\par

Let us now try to understand the behavior of the curve far from the
singularity. This can be analyzed in the regime where $\phi _*/\mu \ll
1$. In this situation, Eq.~(\ref{nnew}) tells us that
\begin{equation}
\label{spectral}
  1-n_{_{\rm S}}\simeq \frac{\mpl^2}{2\pi\mu^2}\, .
\end{equation}
Moreover, since the Lambert function is small for small values of its
argument, this also implies from Eq.~(\ref{phistar}) that [where we
use $W_0(x)\simeq x$]
\begin{equation}
\label{phiCMBSmall}
  \left(\frac{\phi_*}{\mu}\right)^{\!2} \simeq
  \left(\frac{\phi_\mathrm{end}}{\mu}\right)^{\!2}
  \exp\!\left[-\left(\frac{\phi_\mathrm{end}}{\mu}\right)^{\!2}
  -\frac{N_*}{2\pi}\frac{\mpl^2}{\mu^2}\right]\, .
\end{equation}
The previous derivation is consistent as long as the term proportional
to $N_*$ in the argument of the exponential in \eqref{phiCMBSmall} is
large rendering the argument of the Lambert function small. This
implies that $N_*\mpl^2/2\pi\mu^2\gtrsim 1$ and provides the
consistency constraint $N_*(1-n_{_{\rm S}})\gtrsim 1$. This means
that, in order for our approximation to hold, $n_{_{\rm S}}$ cannot be
too close to $1$ which is fine since this is precisely the regime that
we are interested in [as we are trying to approximate $M(n_{_{\rm
S}})$ far from the singularity]. Another way to see the same thing is
to remark that, if $n_{_{\rm S}}$ were too close to $1$, then, from
Eq.~\eqref{spectral}, it would follow that $\mu\gg\mpl$, in
contradiction with the hypothesis that $\phi_*$ is exponentially
damped, see Eq.~(\ref{phiCMBSmall}).

One the other hand, as is apparent from Fig.~\ref{M}, we are not
interested either in large values of $1-n_{_{\rm S}}$ which are,
anyway, observationally excluded. Therefore, a Taylor expansion in
this quantity is still valid in our case. In particular, using
Eq.~(\ref{spectral}), one can express $\varphi _{\rm end}$, hence
$\varphi _*$ and $\epsilon _*$, in terms of $1-n_{_{\rm S}}$
only. Finally, since $V_*\simeq M^4$, expanding everything in
$1-n_{_{\rm S}}$, we can write Eq.~\eqref{multi} as
\begin{equation}
\label{Mapprox}
  \left(\frac{M}{\mpl}\right)^{\!4} \simeq
  \frac{45}{4}\frac{Q_\mathrm{rms-PS}^{\;2}}{T^2} (1-n_{_{\rm
  S}})\e^{-1-N_*(1-n_{_{\rm S}})}\, , 
\end{equation}
and we recover exactly the correct damping behavior observed in
Fig.~\ref{M} (red dashed line).

\par

Usually in the literature~\cite{KM,JMbraz}, in the regime $\mu
/\mpl\ll 1$, one expands Eq.~(\ref{end}) to arrive at the expression
$(\varphi _{\rm end}/\mu)\sim 2\sqrt{\pi }\mu /\mpl$. Then, following
the same steps as before, we end up with
\begin{equation}
\label{Mkjm}
  \left(\frac{M}{\mpl}\right)^{\!4} \simeq
  \frac{45}{2}\frac{Q_\mathrm{rms-PS}^{\;2}}{T^2}
  \e^{-N_*(1-n_{_{\rm S}})}\, ,
\end{equation}
This expression has to be compared with Eq.~(\ref{Mapprox}). The
corresponding curve is represented in Fig.~\ref{M} (green
dashed-dotted line). For $n_{_{\rm S}}-1$ not too close to zero, the
two predictions are in good agreement.

\par

Based on the above considerations, we arbitrarily choose to work with
$n_{_{\rm S}}\simeq 0.93$ (compatible with the WMAP data) for which it
follows that $\mu \sim 1.5 \mpl$. In this case, Fig.~\ref{M} indicates
that
\begin{equation}
  \left(\frac{M}{\mpl}\right)^4\sim 10^{-12}\, .
\end{equation}
This value corresponds to a regime which is compatible with the
various approximations discussed above. We will adopt these values
when plotting the quantum effects that we calculate in the following.

\subsection{Hybrid inflation}

This case is slightly different since hybrid inflation is in fact a
two-fields model with the potential~\cite{hybrid}
\begin{equation}
V\left(\varphi ,\psi \right)=\frac12 m^2\varphi ^2+\frac{\lambda '
}{4}\left(\psi ^2-\Delta ^2\right)^2+\frac{\lambda }{2}\varphi ^2\psi
^2\, ,
\end{equation}
where $\varphi $ is the inflaton and $\psi $ the waterfall field.
$\lambda ' $ and $\lambda $ are two coupling constants. The advantage
of hybrid inflation is that inflation can be realized even if these
coupling constants are of order one. The inflationary valley is given
by $\psi =0$ and, in this case, the potential reduces to a potential
similar to the one given in Eq.~(\ref{potential}) with $a=1$ and
$b=1$, provided we have
\begin{equation}
\label{paramhybrid}
M=\frac{\lambda '{}^{1/4}\Delta }{\sqrt{2}}\, ,\quad \mu
    =\sqrt{\frac{\lambda ' }{2}}\frac{\Delta ^2}{m}\, .
\end{equation}

\par

In the hybrid scenario there are {\it a priori} two mechanisms for
ending inflation. Either inflation stops by instability when the
inflaton reaches a value
\begin{equation}
\label{cri}
\phi _{\rm cri}=\frac{\lambda ' }{\lambda }\Delta\, ,
\end{equation}
where the mass in the direction perpendicular to the inflationary
valley becomes negative or the slow-roll conditions are violated which
happen for
\begin{equation}
\frac{\phi _{\epsilon}}{\mpl}=
\sqrt{\frac{1}{8\pi }\left(1-\frac{8\pi \mu ^2}{\mpl^2}
+\sqrt{1-\frac{16\pi \mu ^2}{\mpl^2}}\right)}\, .
\end{equation}
Let us notice that $\phi _{\epsilon}$ does not exist if $\mu>
\mpl/(4\sqrt{\pi })$. Therefore, the final value of the inflaton,
$\phi _\mathrm{end}$, is the maximum of $\phi _{\rm cri}$ and $\phi
_{\epsilon}$. To decide which mechanism is realized in practice
requires the knowledge of the parameters of the model. Of course, the
parameters of the models must satisfy the CMB
normalization~\cite{hybrid}. The value of the field at horizon exit is
given by an equation very similar to Eq.~(\ref{phistar}), namely
\begin{equation}
  \frac{\phi_*}{\mu} = \sqrt{ W_0\!
  \left\{\!\left(\!\frac{\phi_\mathrm{end}}{\mu}\!\right)^{\!2}\!
  \exp\!\left[\frac{\phi_\mathrm{end}^2+(N_*/2\pi)\mpl^2}{\mu^2}\right]\!
  \right\}}\, .
\end{equation}

\par

Then, one can repeat the algorithm described in the previous
subsection to implement the CMB normalization. Here, we assume that
the coupling constants are $\lambda =\lambda '=1$ and use the
following values
\begin{equation}
\Delta =10^{-4}\mpl\, ,\quad m\simeq 0.7\times 10^{-8}\mpl \,
.
\end{equation}
This implies $\mu \simeq 1.01\mpl $ and $n_{_{\rm S}}\simeq 1.08$, \ie
a blue spectrum as expected for hybrid inflation. Let us remark that
in the above case, inflation ends by instability and is vacuum
dominated. This is the regime of interest for us because it would be
pointless to calculate the quantum effects in the regime where the
field dominates since this case reduces to the case of chaotic
inflation.

\subsection{Running-mass Inflation}

Running mass inflation can be realized in four different
ways~\cite{running} that we now very briefly describe. In the
following, we refer to these four different models as RM1 to RM4. From
the expression of the potential~(\ref{potentialrunning}), it is easy
to see that $\varphi _0$ is an extremum of $V(\varphi )$. This is a
maximum if $c>0$ and a minimum if $c<0$. According to the
classification of Ref.~\cite{running}, ``model 1'' (RM1) corresponds
to the case where $c>0$ and $\varphi_\cl<\varphi_0$. In this case,
$\varphi_\cl $ decreases during inflation. ``Model 2'' (RM2) also
corresponds to $c>0$ but, now, with $\varphi_\cl >\varphi_0$ and
$\varphi_\cl $ increases during inflation. ``Model 3'' (RM3) refers to
the situation where $c<0$, for which $\varphi _0$ is a minimum, and
$\varphi _\cl <\varphi _0$ all the time. In this case, $\varphi_\cl$
increases during inflation. Finally, ``model 4'' (RM4) has $c<0$ and
$\varphi _\cl>\varphi_0$ decreases during inflation.

\par

The values of the free parameters $c$ and $\varphi _0$ are constrained
by the CMB measurements. In Ref.~\cite{running}, these constraints
have been studied for the four models evoked above in the parameter
space $(c,\sigma )$, where $\sigma $ is defined by $\sigma =-c \ln
\left(\varphi _{\rm end}/\varphi _0\right)$, $\varphi _{\rm end}$
being the value at which inflation stops. Using $\sigma $ allows us to
express the spectral index and the running of the spectral index as
\begin{equation}
  n_{_{\rm S}}-1 \simeq -2c
  +2\s\e^{-cN_*}
  =-2c +\frac{1}{c}\frac{\dd n_{_{\rm S}}}{\dd\ln k}\, ,
\end{equation}
where $N_*$ has already been defined before. 

\par

Let us now study how inflation ends in those models. {\it A priori},
the end of inflation is found from the condition $\epsilon =1$.
However, it is easy to see that this cannot be achieved for model 3
[for $V(\varphi _0) \neq 0$] and model 4. In this case, another
mechanism must be advocated, presumably of the hybrid type. For
simplicity, in the following, we focus on models 1 and 2 only. In the
case of model $1$, one can also show that the condition $\epsilon =1$
cannot be satisfied because $\epsilon \rightarrow 0$ when $\varphi
\rightarrow 0$ and is bounded by $c^2 \varphi _0^2/(16\pi e M_{_{\rm
Pl}}^2)\ll 1$ in the interval $[0,\varphi_0]$.  However, the slow-roll
parameter $\eta \equiv M_{_{\rm Pl}}^2V''/V$ (this slow-roll
parameters differs from $\epsilon _2$ introduced above) increases when
$\varphi \rightarrow 0$ and inflation stops when $\eta =1$. In the
case of model 2, the condition $\epsilon =1$ is {\it a priori}
possible but we will see that, in practice, the condition $\eta =1$
occurs earlier and, therefore, controls the end of inflation as for
model 1.

\par

As an illustration of model 1, we work with the values $c=0.06$ and
$\varphi _0/M_{_{\rm Pl}}=10^{-6}$. In this case, the end of inflation
occurs at $\varphi _{\rm end}/M_{_{\rm Pl}}\sim 2\times 10^{-14}$. The
parameter $\sigma $ is given by $\sigma \sim 1.064$.  One can check in
Fig.~3 of Ref.~\cite{running} (left panel) that these values are
compatible with the CMB constraints and, moreover, that they
correspond to a situation where the Planck satellite will measure the
running of the spectral index.

\par

For model 2, the situation is slightly more complicated. If we use the
same parameters $c$ and $\varphi_0$, then inflation ends at $\varphi
_{\rm end}/M_{_{\rm Pl}}\sim 0.6$ which is fine since this value is
smaller than one. However, this leads to $\sigma \sim -0.79$ which,
according to Fig.~4 (left panel) of Ref.~\cite{running} is not
acceptable. Clearly, a way to cure the previous problem is to increase
the value of $\varphi _0$. Therefore, for instance, we could try
$\varphi _0/M_{_{\rm Pl}}=0.1 $ and $c=0.06$. In this case $\varphi
_{\rm end}/M_{_{\rm Pl}}\sim 2.95$ and $\sigma \sim -0.2$. The values
of $c$ and $\sigma $ are now compatible with the CMB constraints but,
of course, the value of $\varphi _{\rm end}>M_{_{\rm Pl}}$ is
problematic. For instance, the simple expression of the number of
e-folds given by Eq.~(\ref{Napproxrun}) would not be valid in this
case because we have assumed $\varphi _\cl /M_{_{\rm Pl}}\ll
1$. Moreover, we should also include higher orders terms in the
potential. For these reasons, and since we only want to illustrate how
quantum effects behave in the running mass scenario with simple
formulas, we will continue to assume that $c=0.06$ and $\varphi
_0/M_{_{\rm Pl}}=10^{-6}$ even for model 2.

\par

Finally, once $c$ and $\varphi _0$ have been chosen, the COBE
normalization fixes the scale $M$ through the relation
\begin{equation}
  \frac{M^4}{\mpl^4} = \frac{45}{4} \frac{Q_\mathrm{rms-PS}^{2}}{T^2}
  \frac{\s^2\varphi_0^2}{M_{_{\rm Pl}}^2}
  \exp\!\left(\!-\frac{\s}{c}\e^{-cN_*}\!-2cN_*\!\right) .
\end{equation}
In the case of model $1$ this gives $M\sim 6.4 \times 10^{-7}\mpl$
while for model $2$ we have $M\sim 1.2 \times 10^{-6}\mpl $, where in 
both cases we have used $N_*=50$.

\par

Having chosen the parameters, one must also specify the initial
conditions. The total number of e-folds is given by
\begin{equation}
N_{_{\rm T}}=\frac{1}{c}\ln \left[\left(\ln \frac{\varphi
    _\ini}{\varphi _0}\right)^{-1}\ln \frac{\varphi _{\rm
    end}}{\varphi _0}\right]\, .
\end{equation}
Regardless of the sign of $c$, one can check that $N_{_{\rm
    T}}>0$. For model $1$, one takes $\varphi
    _\ini=\left(1-10^{-5}\right)\varphi _0$ which gives $N_{_{\rm
    T}}\sim 240$, while for model 2 one chooses $\varphi
    _\ini=\left(1+10^{-3}\right)\varphi _0$ which implies that
    $N_{_{\rm T}}\sim 165$.


\section{Results and Discussion}
\label{Results and Discussion}

Having fixed the values of the parameters for the four models under
consideration, we can now compute the quantum effects. Let us first
apply our formalism to the potential \eqref{potential} and study the
behavior of the fluctuations in the different inflationary
scenarios. In this case, Eq.~(\ref{slowvar}) can be integrated exactly
and straightforward calculations lead to
\begin{equation}
\label{abvar}
  \frac{\mean{\df_1^2}}{\mu^2} = \frac{4b}{3n}\frac{M^4}{\mpl^4}\,
  \frac{I_n}{a+b(\varphi _\cl/\mu)^n} \left(\frac{\varphi_\cl}{\mu
  }\right)^{\!2(n-1)} \, ,
\end{equation}
with 
\begin{align}
  I_n &\equiv 4 \!\int ^{\varphi_\ini/\mu}_{\varphi_\cl /\mu } \kern-1em\dd x\,
  x^{3(1-n)}\left(a  +b x^n\right)^3 \\
  &= P_n\!\left(\frac{\varphi_\ini}{\mu}\right)
  -P_n\!\left(\frac{\varphi _\cl}{\mu}\right)\, , 
\end{align}
the function $P_n$ being given by
\begin{equation}
  P_n(x) = x^4 \bigg(b + \frac{4a\,x^{-3n} }{4-3n}
  - \frac{6 a b\,x^{-2n}}{n-2} - \frac{12a\,x^{-n}}{n-4}\bigg) \,.
\end{equation}
This expression is valid as long as $n \neq 2,4$ or $4/3$. These three
particular cases must be treated separately and the corresponding
function $P$ reads
\begin{align}
  P_2(x) &= -2\frac{a}{x^2} + 6a x^2 +b x^4 + 12 a b \ln x \, , \\
  P_4(x) &= -\frac{a}{2x^8} - 3\frac{ab}{x^4} + b x^4 +12 a \ln x \, ,
  \\ P_{4/3}(x) &= 9 a b x^{4/3}+\frac92 a x^{8/3} + b x^4 + 4a \ln
  x\, .
\end{align}
It is worth noticing that setting $a =0$ and $b=1$ we correctly
recover the result obtained for a simple power law potential
$V(\varphi)\propto\varphi^n$, namely
\begin{equation}
  \frac{\mean{\df_1^2}}{\mu ^2}= \frac{4}{3n}\frac{M^4}{\mpl ^4}
  \left(\frac{\varphi_\cl}{\mu}\right)^{\!n-2}
  \left[\left(\frac{\varphi_\ini}{\mu }\right)^{\!4} -
  \left(\frac{\varphi_\cl}{\mu }\right)^{\!4}\right]  .
\end{equation}
This result was also derived in Ref.~\cite{MM1}.

\par

The next step consists in calculating the correction to the mean
value. Using Eq.~(\ref{slowmean}), this is immediately done once the
variance is known and the result is
\begin{align}
\label{abmean}
  &\frac{\mean{\df_2}}{\mu} =
  Q_n\!\left(\frac{\varphi_\cl}{\mu}\right)\frac{\mean{\df_1^2}}{\mu^2}
  + \frac{4b}{3n}\frac{M^4}{\mpl^4} \notag \\ &\times
  \frac{(\varphi_\cl/\mu)^{n-1}}{\sqrt{a+b(\varphi_\cl/\mu)^n}}
  \!\left[R_n\!\left(\frac{\varphi_\ini}{\mu}\right) -
  R_n\!\left(\frac{\varphi_\cl}{\mu}\right)\right],
\end{align}
where the two functions $Q_n$ and $R_n$ are defined by the following
expressions
\begin{equation}
  Q_n(x)\equiv \frac{2a(n-1) + b(n-2) x^n}{4x (a + b x^n)}\, ,
\end{equation}
and
\begin{equation}
  R_n(x)\equiv x^{2(1-n)}(a+bx^n)^{5/2}\, .
\end{equation}
Let us notice that, for $a=0$, these functions are simple power-laws
and that, in this case, we recover the formulas found in
Ref.~\cite{MM1} for generic monomial potentials.

\par

The last step consists in computing the volume effects. For this
purpose, we have to compute the integral in Eq.~(\ref{voleffects}) for
the potential \eqref{potential}. This leads to
\begin{widetext}
\begin{equation}
  \frac{\mean{\varphi}_{\rm v} - \mean{\varphi}}{\mu} =
  \frac{32\pi}{n^2}\frac{M^4\mu^2}{\mpl^6}
  \left(\frac{\varphi_\cl}{\mu}\right)^{\!n-1} \left\{
  \frac{4}{\sqrt{a+b(\varphi_\cl/\mu)^n}}
  \left[S_n\!\left(\frac{\varphi_\ini}{\mu}\right) -
  S_n\!\left(\frac{\varphi_\cl}{\mu}\right)\right] -
  P_n\!\left(\frac{\varphi_\ini}{\mu}\right) +
  P_n\!\left(\frac{\varphi_\cl}{\mu}\right)\right\},
\end{equation}  
\end{widetext}
where $S_n$ is the function defined by the expression
\begin{align}
  S_n(x) &= \int \dd x \frac{(a+bx^n)^{7/2}}{x^{3(n-1)}} \\
  &= a ^{7/2}\frac{x^{4-3n}}{4-3n}\,
  {}_2F_1\!\!\left(\frac{4-3n}{n}, -\frac{7}{2};
  \frac{4-2n}{n}; -\frac{b}{a}x^n\right) \! \notag , 
\end{align}
for $a\neq 0$. In this expression ${}_2F_1(a,b;c;x)$ is the
hypergeometric function taking real values for values of the argument
$x$ less than one~\cite{Grad}. This is indeed the case since, for
$b=1$, the argument is negative and for $b=-1$ we always have
$(\varphi_\cl/\mu)^n<1$. If $a=0$ (in the large field models case),
then the argument is ill-defined. However, using the relation for the
hypergeometric function of argument $1/x$~\cite{Grad} (that can be
safely applied since, in this case, we have $b=1$) we obtain a
well-defined expression that yields
\begin{equation}
  S_n(x) = 2 \,\frac{x^{4+n/2}}{8+n} \,,
\end{equation}
up to an unimportant additive constant.

\par

\begin{figure*}[t]
  \includegraphics[width=\columnwidth]{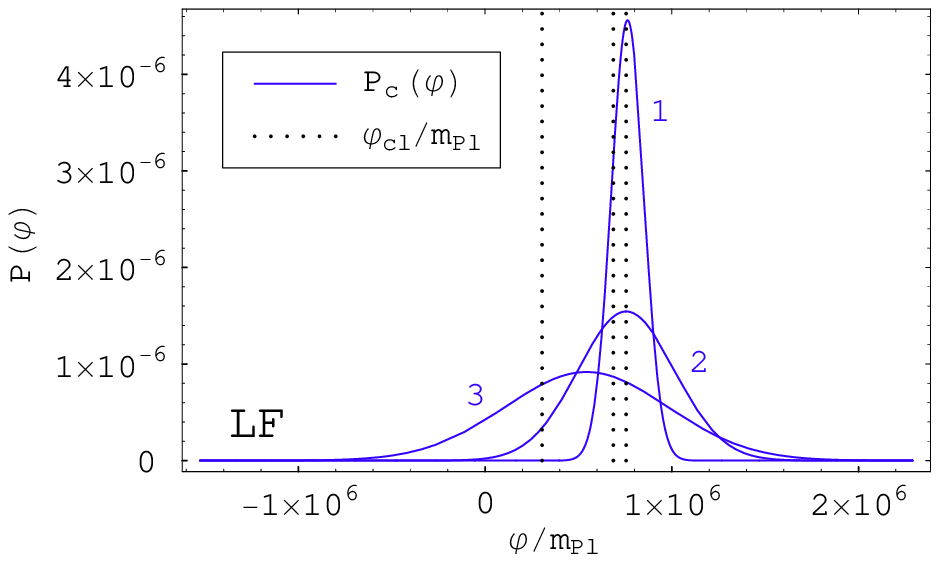}
  \includegraphics[width=\columnwidth]{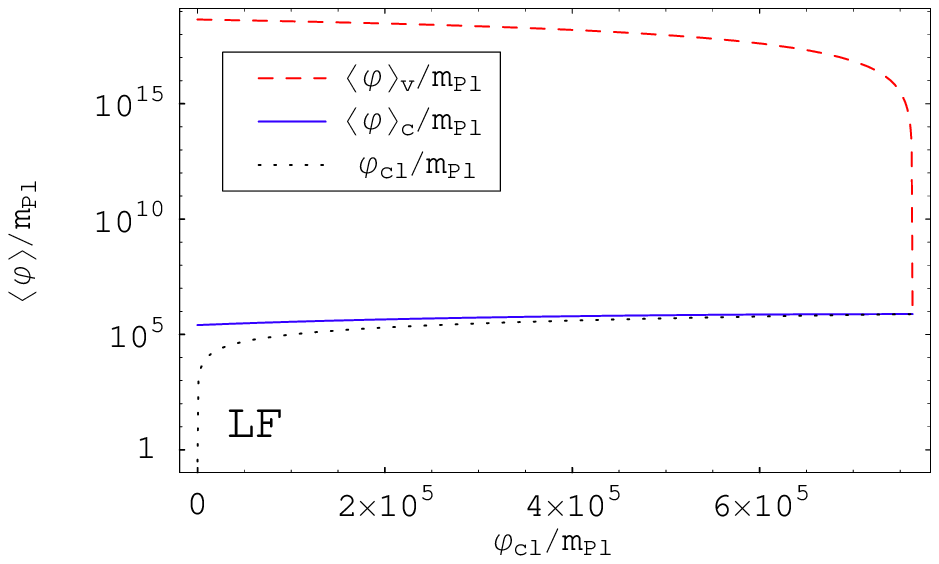}
\\
  \includegraphics[width=\columnwidth]{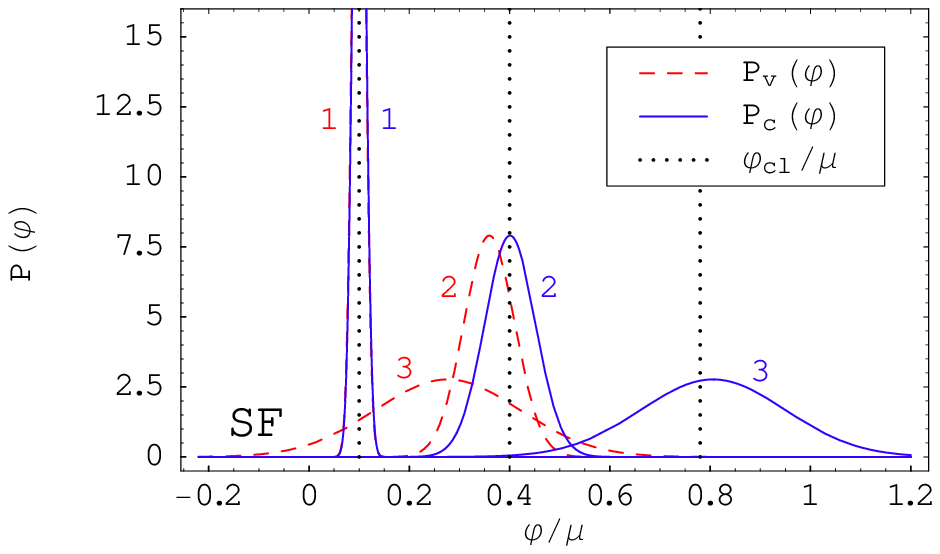}
  \includegraphics[width=\columnwidth]{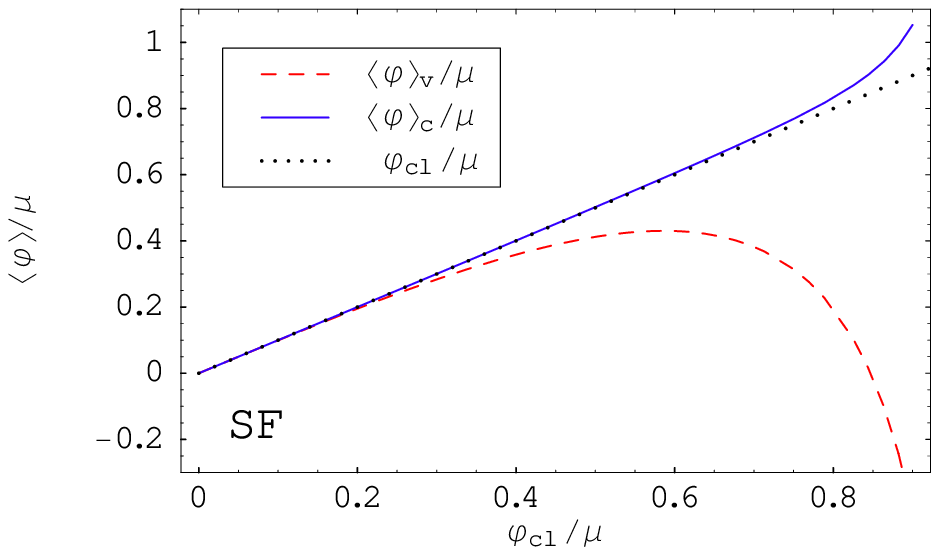}
\caption{\label{largesmall} Evolution of the single-point (solid blue
  line) and volume-weighted (dashed red line) probability
  distributions, $P_\mathrm{c}(\varphi)$ and $P_\mathrm{v}(\varphi)$,
  for the large field model $V\propto\varphi^2$ (LF) and the small
  field model $V\propto 1-(\varphi/\mu)^2$ (SF). The initial values
  are $\varphi_\ini=7.6\times 10^5\mpl$ (corresponding to
  $V_\ini=\mpl^4/2$) for LF and $\varphi_\ini/\mu=10^{-5}$ for SF. The
  initial shape of the two probability density functions is always
  chosen to be $\delta(\varphi-\varphi_\ini)$. The vertical dotted
  black lines represent the location of the classical field. Three
  successive snapshots of the distributions (numbered $1$, $2$ and
  $3$) are shown on the left panels while the evolution of
  $\mean{\varphi}_\mathrm{c}$ and $\mean{\varphi}_\mathrm{v}$ is
  displayed on the right panels. The classical field $\varphi_\cl$
  evolves from the right to the left in LF and from the left to the
  right in SF. In both cases, $P_\mathrm{c}(\varphi)$ rolls down the
  potential, spreads significantly around its mean value and penetrates
  into a classically forbidden region ($\varphi<0$ for LF and
  $\varphi>\mu$ for SF). The quantity $\mean{\varphi}_\mathrm{c}$
  stays ``behind'' the classical value in LF but is ``ahead'' in SF.
  On the other hand, $P_\mathrm{v}(\varphi)$ (not shown in the LF left
  panel) reverses its motion and climbs towards the trans-Planckian
  region (in LF) or towards the maximum of the potential at
  $\varphi=0$ in SF.}
\end{figure*}

Once the variance and the mean value are known, calculating the
probability distribution of the field is straightforward, see
Eq.~(\ref{pc}). Results for large field and small field models are
shown in Fig.~\ref{largesmall}.  For LF, we take an initial condition
$\varphi_\ini $ such that the potential energy is close to the Planck
scale $\mpl^4$. In this case, $P_\mathrm{c}$ (or rather its maximum
$\mean{\varphi}_\mathrm{c}$) slowly rolls down the potential while
remaining ``behind'' the classical solution. At the same time and very
quickly after the initial time, the variance significantly increases,
\ie $P_\mathrm{c}$ strongly spreads, and, as a consequence, the tail
of $P_\mathrm{c}$ penetrates into the region where $\varphi <0$. On
the contrary, $P_\mathrm{v}$, while also spreading a lot, immediately
inverts its motion, starts rolling up the potential and penetrates
into the trans-Planckian regime where $V\gg \mpl^4$. These behaviors
confirm the importance of quantum fluctuations in the LF models but
this also rises the question of the reliability of the solutions
obtained before. Indeed, in the case of $P_\mathrm{c}$, for
$\varphi<\varphi_\mathrm{end}$ (a few $\mpl$'s) the Hubble crossing of
the modes stops and the noise no longer exists. Moreover, the
slow-roll approximation that we have explicitly used in our derivation
of the Langevin equation is no longer valid. With regards to
$P_\mathrm{v}$, in the regime where $V>\mpl^4$, the whole framework of
quantum field theory itself breaks down. These features are also very
sensitive to the initial conditions. For instance, if we have
initially $V(\varphi_\ini)<\mpl^4/10$, then the stochastic deviations
from the classical trajectory are already dramatically reduced.

\par

In the case of SF models similar conclusions hold. If $\varphi_\ini$
is very small (close the maximum of the potential), then
$P_\mathrm{c}$ significantly spreads around its mean value while
rolling down the potential. The only difference is that, now, the peak
of the distribution stays ahead of $\varphi_\cl$. As before, the tail
of the single-point distribution goes to the region where the
slow-roll approximation breaks down, $\varphi>\varphi_\mathrm{end}$
and $\varphi>\mu$. In this case, the previous considerations on the
reliability of the solution still apply. Let us now study the behavior
of $P_\mathrm{v}$. As can be seen in Fig.~\ref{largesmall}, at the
beginning of inflation, the behaviors of $P_\mathrm{c}$ and
$P_\mathrm{v}$ are similar but, after a few e-foldings, $P_\mathrm{v}$
reverses its motion and starts moving back towards the maximum located
at $\varphi=0$. After enough time, most of $P_\mathrm{v}$ is on the
other side of the potential. Diffusion is less crucial than in the LF
models because the energy scales involved are smaller but,
nonetheless, we see that such effects are very important. On the other
hand, the behavior of $P_\mathrm{v}$ is less problematic since the
Langevin equation is fully trustable even if the field is negative
(provided, of course, it is still small in comparison to $\mu $). The
only danger comes from a possible breakdown of the perturbative
expansion as the field goes to a region where the stochastic mean
value is very far from its classical counterpart. In this case, the
quantum effects should in fact be computed as perturbations of a
classical solution living in the region $\varphi <0$. Then, the volume
effects will act the same way as before, that is to say they will push
the corresponding distribution back to the origin, towards the region
$\varphi >0$. Therefore, it seems reasonable to postulate that a
situation of dynamical equilibrium could set up with a stationary
solution for $P_\mathrm{v}$ concentrated around $\varphi =0$. 

\par

Finally, let us notice that, as before, the stochastic behavior of the
field is strongly dependent on the initial conditions. If $\varphi$ is
far from the maximum, then the drift due to the classical term
dominates on the noise-induced diffusion, the effect of which is
therefore no longer important. The dependence on the initial
conditions can be also be checked with the help of Eq.~\eqref{abvar}
in the limit $\varphi_\cl\ll\mu$. In this regime we have $P_2(x)\simeq
-2/x^2$ and, since $\varphi_\ini<\varphi_\cl$, the relative amplitude
of the fluctuations can be written as
\begin{equation}
  \frac{\sqrt{\mean{\df_1^2}}}{\varphi_\cl}\lesssim \sqrt{\frac{4}{3}}
  \left(\frac{M}{\mpl}\right)^{\!2}\frac{\mu}{\varphi_\ini}
  \sim10^{-1}\, ,
\end{equation}
for $M\sim 10^{-3}\mpl $ and $\varphi _\ini\sim 10^{-5}\mu $. This
shows that, in this case, the stochastic fluctuations can represent up
to 10\% of the field value (and they increase in the subsequent
evolution since, in new inflation, different classical solutions
diverge). However, taking another initial condition, for instance
$\varphi_\ini/\mu\simeq 10^{-3}$, would reduce the fluctuations to
only 0.1\% of the classical field. Therefore, we see that there is
indeed a strong dependence on the initial conditions.

\par

Let us now consider hybrid inflation in the vacuum dominated
regime. In this case, the corresponding distributions are so peaked
that they cannot be distinguished from a Dirac distribution. This is
the reason why we have chosen not to include any figure for the hybrid
case. The reason for this behavior can be easily understood if we come
back to Eq.~\eqref{abvar}. In this case,
$\varphi_\ini>\varphi_\cl>\varphi_\mathrm{cri}$ and, therefore, one
has
\begin{equation}
  \frac{\sqrt{\mean{\df_1^2}}}{\varphi_\cl}\sim \sqrt{\frac{4}{3}}
  \left(\frac{M}{\mpl}\right)^{\!2}\frac{\mu}{\varphi_\cl} \, ,
\end{equation}
the upper limit of the above expression being obtained when evaluated
for $\varphi =\varphi _{\rm cri}$. Using Eqs.~(\ref{paramhybrid})
and~(\ref{cri}), one arrives at
\begin{equation} 
\frac{\sqrt{\mean{\df_1^2}}}{\varphi_\cl}\lesssim \frac{\Delta
  ^3}{m\mpl ^2}\sim 10^{-4}\, ,
\end{equation}
for the values of $\Delta $ and $m$ chosen before. Therefore, in
hybrid vacuum dominated models, the fluctuations are less than 0.1\%
of the classical field until the very end of inflation. In addition,
if one uses Eq.~(\ref{abmean}) in the same limit, one sees that the
two terms of that formula exactly cancels out and $\mean{\df_2}\sim
0$. This means that, with a very good accuracy, the distributions are
peaked over the classical value of the field. This explains the
behavior of the distribution functions described before. Moreover,
this conclusion is rather independent of the initial condition,
provided that the latter does not violate the vacuum dominated regime.

\begin{figure*}
  \includegraphics[width=\columnwidth]{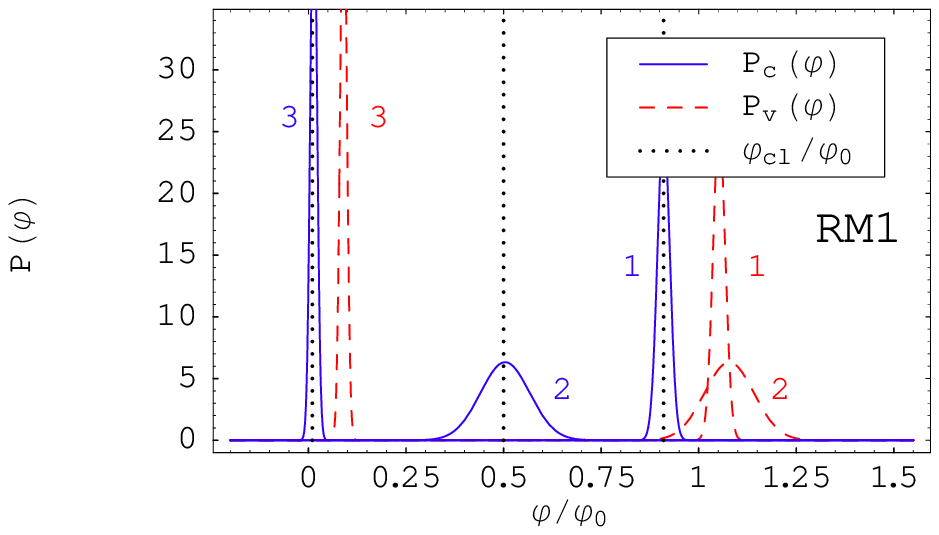}
  \includegraphics[width=\columnwidth]{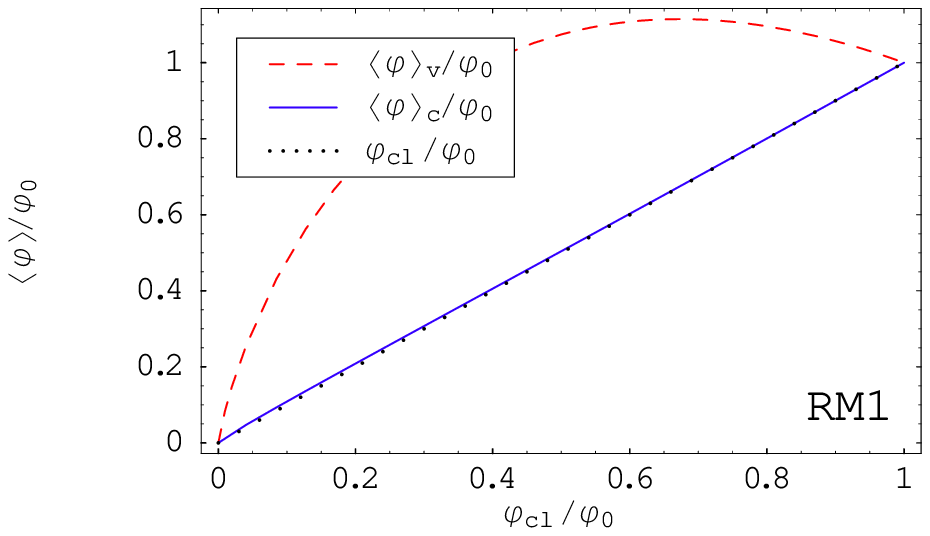} \\
  \includegraphics[width=\columnwidth]{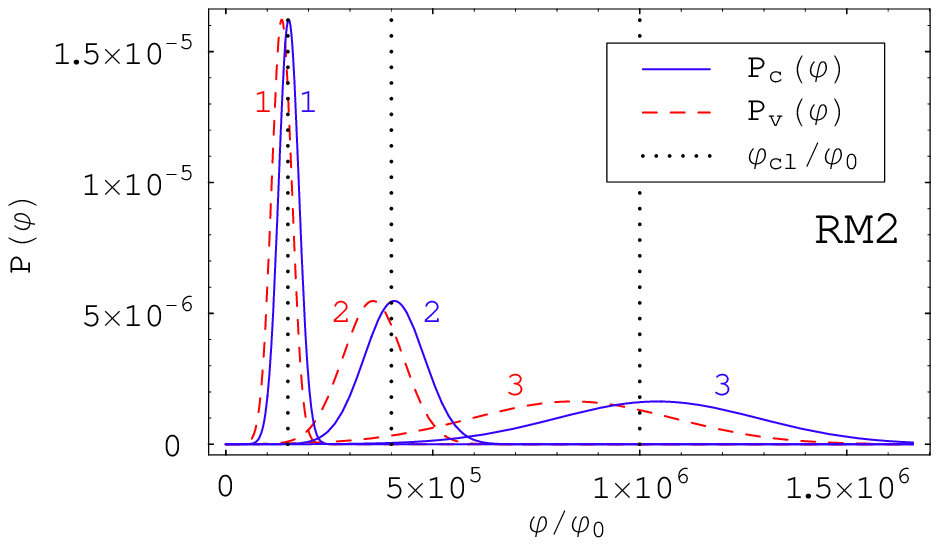}
  \includegraphics[width=\columnwidth]{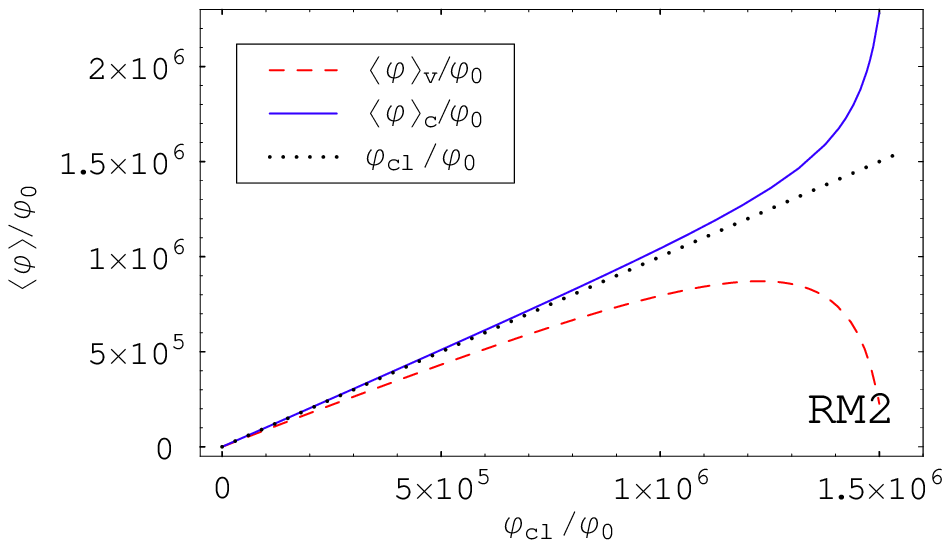}
\caption{\label{RM1and2} Evolution of the single-point (solid blue
  line) and volume-weighted (dashed red line) probability
  distributions for the two different running mass inflation models
  RM1 and RM2. The initial probability density function is
  $\delta(\varphi-\varphi_\ini)$ for the two models. The initial
  values are chosen to be $\varphi_\ini/\varphi_0=1-1.5\times 10^{-5}$
  for RM1 ($\varphi<\varphi_0$, inflation proceeding from the right to
  the left) and $\varphi_\ini/ \varphi_0=1+10^{-3}$ for RM2
  ($\varphi>\varphi_0$, inflation proceeding from the left to the
  right). The vertical lines (dotted black lines) represent the
  location of the classical field. On the two left panels, three
  snapshots of $P_\mathrm{c}(\varphi)$ and $P_\mathrm{v}(\varphi)$
  (numbered from $1$ to $3$) are shown at three different times
  (respectively corresponding for RM1 to $\varphi /\varphi _0=0.9,
  0.5$ and $0.01$) together with the corresponding values of the
  classical field. In the right panels, the evolution of
  $\mean{\varphi}_\mathrm{c}$ and $\mean{\varphi}_\mathrm{v}$ is
  followed until the end of inflation. The physical interpretation of
  these results is discussed in the text.}
\end{figure*}

Let us now investigate the quantum effects for the two different
models of running mass inflation. The variance and the mean value can
be calculated exactly since the corresponding integrals are
feasible. This results in quite complicated formulas which are not
especially illuminating. However, since the vacuum expectation value
of the inflaton is always small with respect to the Planck mass, it is
therefore a good approximation to replace $V'/V$ by $V'/M^4$ as we
have already done before, see Ref.~\cite{running}. Then, the variance
reads
\begin{widetext}
\begin{eqnarray}
  \frac{\mean{\df_1^2}}{\MPl^2} 
  &=& \frac{1}{12\pi ^2 c} \left(\frac{M}{\MPl}\right)^4 
  \left(\frac{\varphi _\cl}{\varphi_0}\right)^2
  \ln ^2 \left(\frac{\varphi_\cl}{\varphi_0}\right) 
  \biggl[\frac12 \left(\frac{\varphi_0}{\varphi_\ini}\right)^2 
  \frac{1}{\ln ^2\left(\varphi _\ini/\varphi _0\right)}
  -\frac12 \left(\frac{\varphi_0}{\varphi_\cl}\right)^2
  \frac{1}{\ln ^2\left(\varphi _\cl/\varphi _0\right)} 
  \nonumber \\
  & & + \left(\frac{\varphi_0}{\varphi_\cl}\right)^2 
  \frac{1}{\ln\left(\varphi_\cl/\varphi _0\right)}
  -\left(\frac{\varphi_0}{\varphi_\ini}\right)^2
  \frac{1}{\ln \left(\varphi_\ini/\varphi_0\right)}
  +2 {\rm Ei}\left(-2 \ln \frac{\varphi_\cl}{\varphi_0}\right)
  -2 {\rm Ei}\left(-2 \ln \frac{\varphi_\ini}{\varphi_0}\right)\biggr]\, .
\end{eqnarray}
>From this result, using Eq.~(\ref{slowmean}), we deduce the expression
of the mean value, namely
\begin{eqnarray}
  \frac{\mean{\df_2}}{M_{_{\rm Pl}}} &=& \frac12\left[\frac{1+\ln
  \left(\varphi _\cl/\varphi _0\right)}{\left(\varphi _\cl/M_{_{\rm
  Pl}}\right)\ln \left(\varphi _\cl/\varphi _0\right)}
  +\frac{c}{2}\frac{\varphi }{M_{_{\rm Pl}}}\ln \frac{\varphi _\cl
  }{\varphi _0}\right]\frac{\mean{\df_1^2}}{M_{_{\rm Pl}}^2}
  +\frac{1}{48\pi ^2c}\left(\frac{M}{M_{_{\rm Pl}}}\right)^4
  \frac{\varphi _\cl}{M_{_{\rm Pl}}}\ln \left(\frac{\varphi
  _\cl}{\varphi _0}\right) \nonumber \\ & &  \times \left[
  \left(\frac{M_{_{\rm Pl}}}{\varphi _\cl}\right)^2 \frac{1}{\ln
  ^2\left(\varphi _\cl/\varphi _0\right)} -\left(\frac{M_{_{\rm
  Pl}}}{\varphi _\ini}\right)^2 \frac{1}{\ln ^2\left(\varphi
  _\ini/\varphi _0\right)}\right]\, .
\end{eqnarray} 
\end{widetext}
Finally, let us compute the volume effects given by
Eq.~\eqref{voleffects}. Following the approximation already used
above, one could take a factor $H$ out of the integral. Then, the
remaining integral, the kernel of which is now $(H/H')^3$, can be
performed explicitly. However, in this case, the result exactly cancel
the second term in Eq.~\eqref{voleffects} and we would obtain
$\mean{\varphi}_{\rm v} - \mean{\varphi}\simeq 0$.  Let us notice that
this is is not a specific feature of the running mass potential but a
general consequence of the approximation used. Therefore, in order to
compute the volume effects, one should evaluate the original
integral. Unfortunately, an exact integration is not possible in this
case. A possible solution is then to expand the integrand in powers of
$c(\varphi_0/\MPl)^2$. At first order, this gives
\begin{widetext}
\begin{eqnarray}
3I^TJ &=&-\frac{7M}{16 \pi ^2c}\frac{M^3}{M_{_{\rm Pl}}^3}
\frac{\varphi _\cl}{M_{_{\rm Pl}}}\ln \frac{\varphi _\cl}{\varphi _0}
\left[1-\frac{c}{2}\left(-\frac12+\ln \frac{\varphi _\cl}{\varphi _0}
\right)\frac{\varphi _\cl^2}{M_{_{\rm Pl}}^2}\right]^{-1/2}
\nonumber \\
& & \times
\left[-\frac{1}{\ln (\varphi _\ini/\varphi _0)}
+\frac{1}{\ln (\varphi _\cl/\varphi _0)}
+\frac{1}{4\ln ^2(\varphi _\ini/\varphi _0)}
-\frac{1}{4\ln ^2(\varphi _\cl/\varphi _0)}\right]\, .
\end{eqnarray}
\end{widetext}
Alternatively, a numerical integration of Eq.~\eqref{voleffects} can
always be done and, in the following, we present results obtained with
this method.

\par

Results for RM1 and RM2 are displayed in Fig.~\ref{RM1and2}. Let us
start with the RM1 model. In this case, $P_\mathrm{c}(\varphi)$
follows very closely the classical solution, the peak of the
distribution being slightly ``behind'' $\varphi _\cl$ (as for
LF). Interestingly enough, if, at the beginning of inflation, the
distribution $P_\mathrm{c}(\varphi)$ starts spreading around its mean
value as it was the case for the LF and SF models then, after some
e-foldings, the variance reaches a maximum and then starts decreasing,
\ie $P_\mathrm{c}(\varphi)$ becomes more and more peaked over the
classical solution. This is a consequence of the fact that the
classical dynamics (which dominates at late times) tends to attract
different solutions contrarily to the SF case where they are instead
pulled apart. The behavior of $P_\mathrm{v}(\varphi)$ is even more
interesting. The volume weighted distribution moves backwards, beyond
the maximum of the potential, into the region classically
corresponding to RM2. Then, it reverses its motion and comes back into
the region corresponding to RM1. This last behavior is probably not
trustable because, as long as $P_\mathrm{v}(\varphi)$ penetrates the
region RM2, the calculations performed before should be modified to
take this situation into account. Let us also notice that this
behavior is not exactly similar to the one observed for the SF model
where the distribution goes to the region $\varphi <0$, see the
discussion above. Indeed, in this last case, whatever the sign of the
field, the model is the same, in particular the value of $M$ remains
unchanged. On the contrary, RM1 and RM2 are really two different
models with two different energy scales. Let us now study RM2. As for
RM1, the peak of $P_\mathrm{c}(\varphi)$ follows the classical
solution with the difference, however, that it is slightly ahead
$\varphi _\cl$ (as for SF) and that the spreading of the distribution
continuously increases. At the beginning of inflation,
$P_\mathrm{v}(\varphi)$ also follows $\varphi _\cl $ but, then, it
changes its motion and moves back towards the region corresponding to
RM1.

\par

One of our main conclusion concerning running mass inflation is that,
if RM1 and RM2 are classically two different models, from a
statistical point of view, it may then be impossible to distinguish
between them. Indeed, regardless of which model we decide to start
with, if the initial conditions are sufficiently close to $\varphi_0$,
then there is a significant probability of diffusing on the other side
of the potential. Moreover, when we start with RM1 (RM2), volume
effects tends to push the evolution towards RM2 (RM1). Therefore, as
it was the case for SF, it seems reasonable to postulate that the
system will settle down at the boundary between the two models, a
situation described by a stationary distribution concentrated around
$\varphi _0$. Such a situation typically leads to the self-reproducing
regime. 

\par

Finally, another important feature of the running mass potential is
that quantum effects can strongly modify the classical evolution even
if the energy scale involved is far below the Planck energy (this is
also the case for new inflation). This reinforces the fact that the
connection between the importance of the quantum effects on one hand
and the fact that $V$ is close to the Planck energy on the other hand
is very specific to the large field monomial potentials used in
chaotic inflation. In fact, it is clear that quantum fluctuations are
important whenever the classical contribution to the motion of the
field is suppressed. In chaotic inflation, this happens close the
Planck scale because of the large friction term but this also happens
near the maximum of the potential (and far from the Planck scale) in
new and running mass inflation models because of the smallness of
$V'(\varphi)$.

\par

We end this section by a slightly different discussion. In
Ref.~\cite{GT}, eternal inflation is studied using the Langevin
equation for the model $V(\varphi )\propto \varphi ^4$. The main
argument presented for considering this simple model is that
analytical progress may be made in this case. Using the method
presented in our article, one can in fact consider a much larger
variety of models and one is not restricted to the simple chaotic
quartic model. In order to illustrate this claim we consider the
calculation, presented in Ref.~\cite{GT}, of the value of the inflaton
at which the self-reproducing behavior becomes possible, not only for
the potential $\varphi ^4$ but for the general case $\varphi ^n$ (in
fact, one could even reproduce this calculation for new inflation,
running mass inflation and so on).

\par

In Ref.~\cite{GT}, the main idea is to evaluate the mean value of the
number of e-folds is given by
\begin{equation}
  \mean{N} =\int _0^T \dd t \mean{H} = -\frac{\kappa}{2} \int
  _{\varphi _\ini}^{\varphi(T)}\dd \psi \frac{\mean{H}}{H'_\cl}\, .
\end{equation}
This expression can easily be evaluated for any potential using our
perturbative treatment. For large fields models, straightforward
calculations lead to
\begin{equation}
\mean{N} =N_{_{\rm T}}^{\rm cl}\left[1+\frac{4(3n+4)M^4}{3n(n+2)\mpl
^4} \left(\frac{\varphi _{\ini}}{\mpl}\right)^n+\cdots \right]\, .
\end{equation}
The breakdown of this expansion signals the beginning of the
self-reproducing regime. Using the fact that $N_{_{\rm T}}^{\rm
cl}\simeq 4\pi (\varphi _\ini/\mpl)^2/n$, one easily sees that this
happens when the initial value of the field is
\begin{equation}
\frac{\varphi _\ini}{\mpl}\sim \lambda _n^{-1/(n+2)}\, ,
\end{equation}
where the coupling constant $\lambda _n$ is defined such that
$M^4/\mpl^n=\lambda _n/\mpl ^{n-4}$. For $n=4$ one recovers the
condition found in Ref.~\cite{GT}. Our method allows us to obtain this
condition for any potential and there is no need to assume a quartic
potential to perform this calculation explicitly. Incidentally, the
above condition was also found previously for instance in
Ref.~\cite{N} [see Eq.~(3.31)].

\vspace*{0pt}

\section{Conclusions}

We now quickly summarize what are the new results obtained in this
article. First, we have presented a perturbative method, used for the
first time in Ref.~\cite{GLMM}, for solving the Langevin equation of
stochastic inflation with, and this is crucial for the present
article, the backreaction taken into account. We have compared this
formalism with the other methods already known in the literature and
have argued that it is more powerful because the approximation is made
directly in the Langevin equation rather than in its solution. In
particular, we were able to provide a general second order expression
for the probability distribution of the field with or without the
volume effects taken into account. Our expression only requires the
calculation of one quadrature (two if the volume effects are
considered) which, for most of the inflationary scenarios, can be
performed explicitly. Second, we have applied this method to various
models of inflation. This has allowed us to compute the quantum
effects with backreaction for chaotic, new, hybrid and running mass
inflation. To our knowledge, in the case of the last three models,
this is the first time that such a calculation is done (the
backreaction being taken into account). Third, we have discussed the
impact of the stochastic effects on these inflationary scenarios. For
instance, in the case of running mass inflation, it was shown that the
quantum effects blur the distinction between the various running mass
inflationary models and that the self-reproducing regime is likely to
be important.

\par

Finally, an important advantage of the method presented here is that
its accuracy and domain of validity can be evaluated in details. This
will be the subject of a forthcoming paper~\cite{MM3}.

\section{Acknowledgments}

M.~M. would like to thank the Universit\`a degli Studi di Milano and
the Institut d'Astrophysique de Paris (IAP) where part of this work
has been done. We would like to thank Patrick Peter for careful
reading of the manuscript. It is a pleasure to thank Andr\'e
Grodemouge for useful comments.

The work of M.~M. is supported by the National Science Foundation 
under Grant Nos. PHY-0071512 and PHY-0455649, and with grant support 
from the US Navy, Office of Naval Research, Grant Nos. N00014-03-1-0639 
and N00014-04-1-0336, Quantum Optics Initiative.

\appendix

\onecolumngrid
\section{General Definitions}
\label{appvolume}

In this short Appendix, we give the precise definitions of the
quantities that we have been using in order to establish the expression of
the probability density function in Sec.~\ref{Solving stochastic
inflation}~C. If we define a ``two-step'' function 
$\vartheta(t_\ini<\t<t)\equiv\vartheta(t-\t)\vartheta(\t-t_\ini)$
(whose value is 1 if the inequality is true and 0 otherwise),
the expression of the ``vector'' $J$ is 
\begin{equation}
  J_{t,t_\ini}(\t) = 
  \vartheta(t_\ini<\t<t) \frac{H'_\cl(t)}{2\pi}
  \frac{H^{3/2}_\cl(\tau)}{H'_\cl(\tau)}
\end{equation}
while the ``matrix''{\bf A} can be written as 
\begin{align}
  \mathbf{A}_{t,t_\ini}(\s_1,\s_2) = \frac{H'_\cl(t)}{4\pi}
  \left[3 \vartheta(t_\ini<\s_1<t)
  \sqrt{H_\cl(\s_1)} J_{\s_1,t_\ini}(\s_2)
  -\frac{\mpl^2}{2}\!\int_{t_\ini}^t \!\dd \t
  \frac{H'''_\cl(\tau)}{H'_\cl(\tau)}
  J_{\t,t_\ini}(\s_1)J_{\t,t_\ini}(\s_2)
  \right] .
\end{align}
Finally, we also provide the definitions of quantities that have been
using in Sec.~\ref{Solving stochastic inflation}~D where the volume
effects have been estimated. In particular, the ``vector'' $I$ can be
expressed as
\begin{equation}
  I_{t,t_\ini}(\s)=3\!\int_{t_\ini}^t\dd\t H'_\cl(\t) J_{\t,t_\ini}(\s)\, ,
\end{equation}
and the ``matrix'' ${\bf B}$ is defined by the following formula
\begin{equation}
  \mathbf{B}_{t,t_\ini}(\s_1,\s_2)= \int_{t_\ini}^t\dd\t
  \left[H'_\cl(\t)\mathbf{A}_{\t,t_\ini}(\s_1,\s_2) +
  \frac{H''_\cl(\t)}{2}J_{\t,t_\ini}(\s_1)J_{\t,t_\ini}(\s_2)\right] \, .
\end{equation}

\twocolumngrid

\end{document}